# Heterophase fluctuations at the gas-liquid phase transition

A.S. Bakai[a]

*NSC Kharkiv Institute of Physics&Technology, Kharkiv 61108, Ukraine*

The problem of the gas-liquid heterophase fluctuations of a fluid within the critical and supercritical regions is revisited. To describe the thermodynamics and structure of the heterophase fluid, the mesoscopic equation of state is deduced. The mesoscopic scale is connected with the sizes of the mixed bubbles and droplets (*g*-fluctuons and *l*-fluctuons). The two-component order parameter of the heterophase fluid describes the gas-like and liquid-like fractions. Its interconnection with the van der Waals order parameter is ascertained. The non-symmetric Ising-type equation of state of the heterophase fluid is analyzed using the standard methods. The manifestations of the mesoscopic effects in the structural and thermodynamic properties of the fluid are investigated. The domain of applicability of a classical model is found using the mesoscopic Ornstein-Zernike equation. The asymmetry and non-analyticity of the diameter of the phase coexistence curve is researched. It is shown that that the critical non-analyticity of the diameter is induced by the mesoscopic heterophase fluctuations. The percolation thresholds of the droplets and bubbles in the critical and supercritical fluid are determined. On the phase diagrams they bound the domains of applicability of the droplet model.

## I. INTRODUCTION

Since the long-range order (LRO) is absent in the amorphous matter, its structure is specified by the short-range order (SRO). The heterophase fluctuations (HPF) produce the SRO heterogeneity.

The HPF are transient nuclei of a new phase within the matrix phase. The term "new phase" means that the nuclei have SRO different than that within the surrounding matrix. Unlike the homophase (perturbative) fluctuations (e.g., elementary excitations), the heterophase fluctuations are non-perturbative, i.e. they can not be considered as small deviations of the physical quantities from their equilibrium values. The states containing statistically significant amount of the HPF are heterophase.

By definition, the gaseous bubbles formed within the liquid and the liquid-like droplets formed within the gas are the HPF. The role of the HPF near the gas-liquid critical point has been a subject of considerations and discussions for a long time. The present paper is devoted to this problem.

---

[a] Electronic mail: bakai@kipt.kharkov.ua

The paradigmatic theory of the gas-liquid critical point was formulated by van der Waals.[1] In spite of the triumph, van der Waals was not satisfied with the consistency of his theory predictions with experimental data. He looked for a reasonable explanation and intended to remove the discrepancy. Because of this, a considerable part of his Nobel lecture[2] was devoted to the question of why the discrepancy occurs and how to resolve this difficulty. Van der Waals noted that the interactions of molecules within the gas become much more significant near the gas-liquid critical point and this causes the formation of "pseudo-associates". The pseudo-associates were described only qualitatively. They can be conceived as the short-lived dense molecular clusters. Van der Waals soundly asserted that the number of pseudo-associates in a gas increases with the decrease of the temperature and/or volume. Evidently, they are non-perturbative fluctuations and can be treated as the HPF in the form of the short-leaving liquid-like droplets in a gas. Evidently, the pseudo-associates generate the SRO heterogeneities. Accepting this concept, one can conclude that the mesoscopic parameters attributed to the pseudo-associates have to be included by some means in the improved theory of the gas-liquid critical point.[b]

Variety of theoretical models of the HPF, starting from the basic Frenkel papers,[4,5] is mostly devoted to the description of the heterophase states far from the critical point (see, e.g., the review paper[6] and the references therein). In this case a comparatively small amount of the matter belongs to the HPF and the HPF can be considered as isolated, non-interacting entities. However, this is not the case within the critical region where, as it was shown,[7,8] the HPF are so strong that the fluid is a two-phase "mixture" with comparable fractions of both phases.

Evidently, the HPF interaction can not be ignored when describing an essentially heterophase state. Therefore, we strive to develop a theoretical model which allows to describe the HPF effects beyond the droplet model used in Refs. 7,8.

The theoretical models of the critical phenomena (see, e.g., books[9-14] and the references therein) are mainly focused on the long-range correlations of the order parameter fluctuations near the critical point. The universality of the critical behavior of different systems shows that the SRO, which is not a universal property, plays a secondary role in the criticality. The success of Landau's theory of the critical point,[9] in which the SRO features can reveal themselves just in the phenomenologic coefficients of the effective Hamiltonian, proves that this statement is correct. For this reason the nonuniversal behavior of the asymmetry of fluid criticality (see Refs. 15-20 and the references therein) enforces one to assume that it is in some way connected with the SRO heterogeneity generated by the HPF near the gas-liquid critical point; and that the mesoscopic SRO heterogeneity has to be explicitly accounted for in the effective Hamiltonian in order to describe the asymmetry effects.

---

[b] Van der Waals was not the first who assumed that the fluid is heterophase near the critical point but his idea is a clue to deducing the mesoscopic theory of critical point. Levelt Sengers has publshed circumstantial retrospective review of the experimental and theoretical papers devoted to the critical point existence and its nature.[3]

The direct correlation function introduced by Ornstein and Zernike[21] can be used to identify the SRO out of the critical region. The Ornstein-Zernike (OZ) integral equation,[21] formulated considering the critical density fluctuations of the fluid,

$$\Gamma(\vec{r}) = \gamma(\vec{r}) + \rho \int \gamma(\vec{r}')\Gamma(\vec{r} - \vec{r}')d^3\vec{r}', \tag{1}$$

interconnects the direct and total correlation functions, $\gamma(\vec{r})$ and $\Gamma(\vec{r})$, respectively. Here $\rho$ is the density.

Ornstein and Zernike also formulated the OZ hypothesis which allows to parameterize the critical long-wave part of the $\gamma(r)$. Namely, they assumed that the Fourier transform of the direct correlation function, $\hat{\gamma}(q)$, can be presented as a Taylor series for $q \to 0$ and $T \to T_c$ ($T_c$ is the critical temperature),

$$\hat{\gamma}(q) = \gamma_0 - \rho^{-1}R^2 q^2 + ..., \tag{2}$$

$$R^2 = \int r^2 \gamma(r) d\vec{r}, \tag{3}$$

where $R$ is the OZ radius. It is the correlation length of the direct correlation function.

The structure factor, $S(q)$, and the Fourier transforms of the $\gamma(\vec{r})$ and $\Gamma(\vec{r})$ are connected by the following relation

$$\frac{1}{\rho}S(q) = 1 + \rho\hat{\Gamma}(q) = \frac{1}{1 - \rho\hat{\gamma}(q)}. \tag{4}$$

At the critical point, at $T = T_c$, $P = P_c$ ($P_c$ is the critical pressure), $\gamma_0$ is equal to the reciprocal density,

$$\gamma_0 = \rho^{-1} \tag{5}$$

The equation (4) determines the well known OZ spectrum of the critical fluctuations.

The OZ hypothesis (2) allows one to close the OZ equation at small $q$ and to determine the spectrum of the classic long-range density fluctuations near the critical point. Other closures of the OZ equation (see e.g.,[22,23]) associate $\gamma(r)$ with the pair interaction potential of the molecules.

The classic theory of the critical phenomena fails within the fluctuation range near $T_c$. The fluctuation theory of the criticality is expounded, e.g., in[9-14]. Considering the fluid criticality in the scaling approach, Fisher checked the applicability of the OZ hypothesis assuming that $\gamma(\vec{r})$ is negligible beyond the range of the pair molecular interaction.[23] He has shown that the substitution of the fluctuation correlation function $\hat{\Gamma}(q)$ into the relations (2)-(4) gives an infinite value of $R$ at $T = T_c$. The consideration of the applicability range of the equations (2)-(4) for describing the heterophase fluid is an issue of special interest which is discussed below.

Above $T_c$, even beyond the fluctuation domain, the fluid remains mesoscopically heterophase. The gas-like state, possessing low density, transforms into the liquid-like state at the densification under pressure due to the increase of the liquid-like fraction. The description of these transient states is one of the goals of the theory of gas-liquid HPF.

The present paper is devoted to the phenomenological theory of the gas-liquid critical point taking into account the HPF. In Section II the general properties of the gas-liquid HPF are described. The basic phenomenological equations of the HPF model (HPFM) are formulated in Section 3. Their solutions and sequences are considered in Sections II-VII. The attention is paid to the manifestations of the SRO heterogeneity in the structural and thermodynamic properties of a fluid and to the asymmetry of the fluid criticality. Sections VIII and IX are devoted to the discussions and summary, respectively. Some calculations are placed in the Appendixes.

## II. HPF AND THE GAS-LIKE AND LIQUID-LIKE FLUCTUONS

### A. Frenkel's theory of the pretransition phenomena and Fisher's droplet model

The Frenkel and Fisher phenomenological droplet models of the HPF are briefly stated below as the referent ones. Their applicability domain is determined in Section VIII.

The Frenkel model describes the non-interacting droplets in a gas and bubbles in a liquid near the gas-liquid coexistence temperature far from the critical point at $T < T_c$.[4,5] The free energy of formation of the spherical gas bubble in a liquid or liquid droplet in a gas is

$$g_i(r,T) = \frac{4\pi}{3}\left(\frac{r}{a_i}\right)^3 (\mu_k - \mu_i) + 4\pi r^2 \sigma = \frac{4\pi}{3}\left(\frac{r}{a_i}\right)^3 (s_i - s_k)(T_e - T) + 4\pi r^2 \sigma \quad . \quad (6)$$

$$i,k = g,l; \quad i \neq k; \quad T < T_c$$

The subscripts "g" and "l" denote the quantities related to the gas and liquid, respectively; $\mu_g, s_g$ and $\mu_l, s_l$ is the free energy and entropy per molecule of the gas and liquid; $\sigma$ is the interfacial free energy; $a$ is the mean distance between molecules; $r$ is the embryo radius. The dependence of the thermodynamic quantities on $T$ and $P$ is not explicitly indicated in (6). The phase coexistence temperature, $T_e(P)$, at fixed $P$ is determined by the Gibbs equation

$$\mu_l(P,T) = \mu_g(P,T)\big|_{T=T_e} . \quad (7)$$

Frenkel assumed that the bubbles and droplets are large, i.e. $r \gg a$. Frenkel size distribution of droplets and bubbles is

$$f_i(r,T) \sim \exp\left[-g_i(r,T)/T\right] \tag{8}$$

Hereafter the Boltzmann constant is put to be equal to 1.

As it follows from (5), as $T \to T_e$

$$g_l(r,T) = g_g(r,T) = 4\pi r^2 \sigma \tag{9}$$

With deviation from the phase coexistence curve both $g_l(r,T)$ and $g_g(r,T)$ increase $\sim |T - T_e|$ due to the contribution of the first term in the right-hand side of (6). Therefore, the HPF effects are essential just in the close vicinity of the phase coexistence temperature. For this reason Frenkel called them "pretransitional effects".

A phenomenological droplet model of the liquid HPF in a gas was used by Fisher[7] to describe the gas-liquid criticality. In this model the droplets are non-interacting but, unlike the Frenkel model, the critical region is included into consideration. The location of the critical point on the phase coexistence curve is determined as the temperature at which the interfacial free energy $\sigma$ becomes equal to 0. It is seen from (6)-(9) that due to this condition $f_l(r,T) \to 1$ as $T \to T_c$ independently of $l$. This unreasonable result appears due to neglecting the interaction of droplets. To overcome this difficulty Fisher introduced the new critical exponent, $\tau$, as follows

$$f_l(k,T) \sim k^{-\tau} \exp\left[-g_l(k,T)/T\right]. \tag{10}$$

here $k$ is the number of molecules within the droplet, the exponent $\tau$ is connected with the critical exponents by the scaling relations.[7] The analysis of the experimental and computer simulation data gives the numerical value $\tau \approx 2.25$.

Fisher did not account for the nucleation of the gas-like HPF within the droplets. A "symmetrized" version of the droplet model is proposed in Ref. 7. An important conclusion which follows from the articles 7,8 is that in the vicinity of critical point the fluid is essentially heterophase and it has to be considered as a "mixture" of bubbles and droplets with comparable fractions of both. A corollary is that the SRO is essentially inhomogeneous in the critical region.

The heterophase patterns of a liquid, gas and fluid in the vicinity of the phase coexistence curve below and within the critical region are schematically shown in Fig.1. The Frenkel model of the heterophase fluctuations describes the states depicted in Fig. 1a and Fig. 1b. The consideration of the essentially heterophase states (Fig. 1c) is an issue of this communication.

To explicitly take into account the heterogeneities of the mesoscopic scale, which appear due to the HPF, the mean density (or specific volume) can not be taken as a proper order parameter. The fractions of coexisting phases, as the functions of $P$ and $T$, are usually taken as the structural and thermodynamic characteristics of the state in theoretical models of the heterophase states.[5]

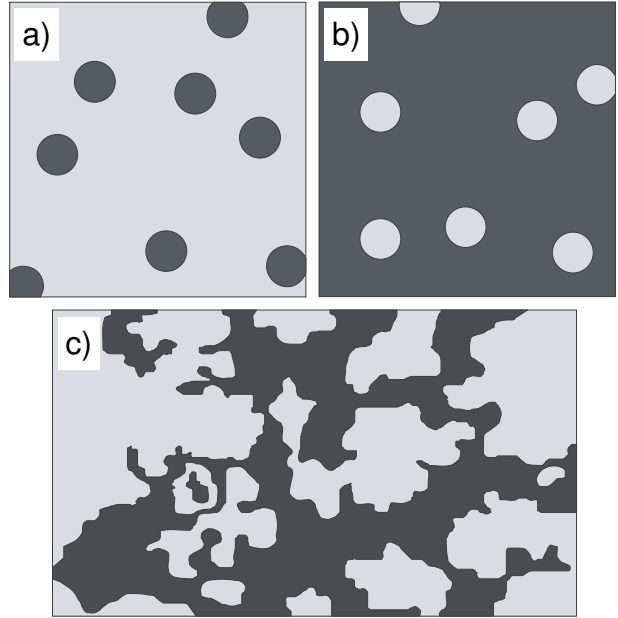

Fig. 1. The schematic patterns of the heterophase states. a) Gas bubbles within the liquid and b) liquid droplets within the gas near the coexistence curve below the critical region. c) The heterophase state near $T_c$; gas-like and liquid-like domains are shown.

A progress in the development of the theory of HPF with multiple types of SRO was made earlier [24-28] while considering the thermodynamics of the supercooled liquids and the liquid-glass transition. The fractions of molecular clusters possessing different SRO are taken as the components of the order parameter. The thermodynamic characteristics and interactions of these clusters determine the structure and thermodynamics of the heterophase state. The liquid-glass transition is in a way associated with the "second critical point" and with the critical phenomena. Therefore, the physics of the heterophase states at the glass-liquid polymorphous transformation has a lot in common with the physics of the gas-liquid transformation. However, the developed model of the liquid-solid HPF can not be directly used to describe the gas-liquid heterophase states. The reason for this is the essential difference between the features of the liquid-solid and gas-liquid phase states. Besides, the supercooled liquid is metastable or unstable. Therefore, at the glass transition both the ergodicity and stability of the system are violated while the gas-liquid transformation can be realized as the evolution of stable states due to shortness of the equilibration time. Thus, the model of HPF developed previously[24-28] has to be modified to describe the gas-liquid criticality.

**B. The gas-like and liquid-like fluctuons**

Let us consider the applicability domain of Eq. (6) more thoroughly. It is correct for large droplets and bubbles, i.e. at sizes $r \gg a$. The lower bound of $r$, at which this equation still makes sense, can be estimated as follows. On one hand, $r$ can not be less than the length $R$ (3). On the other hand, the density fluctuations within the HPF can not exceed the mean density value. These conditions are necessary to determine the HPF of the minimal size.

The bubble of volume $V \sim r^3$ contains $k_g \sim V/v_g$ molecules, $v_g$ is the specific volume of a gas, $k_g$ is the number of molecules within the bubble. The volume fluctuations depend on $V$, $k_g$ and $T$ as follows [8]

$$\left\langle (\Delta v)^2 \right\rangle = \frac{vT\kappa_T}{k_g}, \tag{11}$$

here $\kappa_T$ is the compressibility at constant $T$. For a gas $\kappa_T \sim 1/P$.

For the gaseous phase identity we have to demand

$$\left\langle (\Delta v)^2 \right\rangle \ll v^2. \tag{12}$$

Taking into account that $PV \sim kT$, we have that at $k_g \gg 1$ the specific volume fluctuation within the bubble is small compared with its mean value. The minimal bubble size is $r_{0g} \sim (k_g v_g)^{1/3} = a_g k_g^{1/3}$, $v_g$ is the specific volume of a gas. Since the gas density is comparatively low, just a few molecules can be found in a gas within the range $r \sim R \sim a$. Therefore, $r_{0g} = a k_g^{1/3} > R$.

Considering the liquid droplet in a gas, we have to put $r > r_{0l} \approx R$. It is worth noting that the condition (12) is fulfilled at a large number of molecules within the droplet, i.e. at $k_l \gg 1$, because $\kappa_T$ of a liquid beyond the critical region is much smaller compared with that of gas. Taking $R$ as the minimal radius of the droplet, we can estimate the number of molecules within it. In a cube of linear size $2R$

$$k_l \approx (2R/a)^3. \tag{13}$$

It is apparent that $r_{0l}$ and $r_{0g}$ are the quantities of the same order and that

$$\left| \Delta k / \overline{k} \right| \sim 1, \quad \Delta k = k_l - k_g; \quad \overline{k} = (k_l + k_g)/2. \tag{14}$$

Thus, one has to conclude that in the Frenkel distribution (8) the size of the droplet is bounded below by $r_{0l}$, and the bubble size is bounded below by $r_{0g}$. Under this condition the mean radii of the droplets and bubbles are

$$\left\langle r_l \right\rangle = r_{0l}\left(1 + \frac{T}{8\pi r_{0l}^2 \sigma}\right) \approx r_{0l} \approx R \tag{15}$$

and

$$\langle r_g \rangle = r_{0g}\left(1 + \frac{T}{8\pi r_{0g}^2 \sigma}\right) \approx r_{0g} > R, \qquad (16)$$

respectively.

The obtained estimations characterize the sizes of the fluctuons and the number of molecules per fluctuon[c].

## III. MODEL OF THE GAS-LIQUID HPF

Let us introduce the liquid-like and gas-like fluctuons ($l$- and $g$-fluctuons) as species of the liquid-like and gas-like fractions consisting of $k_l$ and $k_g$ molecules, respectively. Then the total number of the fluctuons, $N_f$, depends on the total numbers of molecules, $N$, and their fractions as follows.

The mean concentrations of the molecules belonging to the liquid-like and gas-like fractions are

$$c_l = \frac{N_l}{N}; \quad \text{and} \quad c_g = \frac{N_g}{N} = \frac{N - N_l}{N} = (1 - c_l), \qquad (17)$$

where $N_l$ and $N_g$ are respectively the number of molecules belonging to the liquid-like and gas-like fraction. Since the numbers of the $l$- and $g$-fluctuons are equal to

$$N_{f,l} = \frac{N_l}{k_l} = N\frac{c_l}{k_l} \text{ and } N_{f,g} = N\frac{c_g}{k_g}, \qquad (18)$$

the total number of the fluctuons is equal to

$$N_f = N_{f,l} + N_{f,g} = \frac{N}{k_g + \Delta k \sigma_l}. \qquad (19)$$

Here

$$\sigma_l = \frac{N_{f,l}}{N_f} = 1 - \frac{N_{f,g}}{N_f} \equiv 1 - \sigma_g \qquad (20)$$

is the fraction of the $l-$fluctuons and $\sigma_g$ is the fraction of the $g-$fluctuons.

It is convenient to introduce the variable $\alpha = \sigma_l - 1/2$. Hence,

---

[c]In the HPF model[25-29] used to describe the glass transition it is assumed that sizes of the solid-like and liquid-like fluctuons are nearly equal because the difference of the densities of the solid and liquid phases is relatively small. Evidently this assumption fails at consideration of the gas-liquid heterophase fluctuations.

$$\sigma_l = \frac{1}{2} + \alpha; \quad \sigma_g = \frac{1}{2} - \alpha \tag{21}$$

Let us denote by $g_2^0 > 0$ the free energy of the pair interaction of the $l$- and $g$-fluctuons. It is proportional to the $l$-$g$ pair interface and to the specific interfacial free energy which appears due to the difference of fluctuons densities.

The phenomenological free energy of the interacting fluctuons can be presented as follows,

$$G(P,T) = G_m(P,T;c_l(x)) + G_f(P,T;\sigma_l(x)), \tag{22}$$

$$G_m(P,T;c_l(x)) = N\int \left[ c_l(x)\mu_l(P,T) + c_g(x)\mu_g(P,T) \right] d^3x, \tag{23}$$

$$c_l(x) + c_g(x) = 1, \tag{23a}$$

$$G_f(P,T;\sigma_l(x)) = \int N_f(x) \left[ \sigma_l(x)\sigma_g(x)g_2 + T\left( \sigma_l(x)\ln\sigma_l(x) + \sigma_g(x)\ln\sigma_g(x) \right) \right] d^3x, \tag{24}$$

$$\sigma_l(x) + \sigma_g(x) = 1. \tag{24a}$$

The fields $c_l(x)$ and $c_g(x)$ describe the local densities of molecules belonging to the $l$- and $g$-fractions; $\sigma_l(x)$ and $\sigma_g(x)$ are the densities of the fluctuons; $g_2 = z_f g_2^0$, $z_f$ is the mean coordination number of the fluctuons.

$G_m(P,T;c_l(x))$ is the free energy of the liquid-like and gas-like fractions not accounting for their interaction. It is symmetric at the transformation $c_l(x) \leftrightarrow c_g(x)$.

$G_f(P,T;\sigma_l(x))$ describes the free energy of the interacting fluctuons. The expression in square brackets is symmetric at the transformation $\sigma_l(x) \leftrightarrow \sigma_g(x)$. Its last term describes the entropy of fluctuons mixing.

The parameters $c_l$ and $\sigma_l$ determined by the relations (17) and (20) are the mean values of the fields $c_l(x)$ and $\sigma_l(x)$: $c_l = \langle c_l(x) \rangle$; $\sigma_l = \langle \sigma_l(x) \rangle$ ($\langle ... \rangle$ means the spatial averaging). Besides,

$$c_l(x) = \frac{k_l \sigma_l(x)}{k_g + \Delta k \sigma_l(x)} = \left( \frac{2+\varsigma}{4} \right) \frac{1+2\alpha(x)}{1+\varsigma\alpha(x)}, \tag{25}$$

$$\varsigma = \frac{\Delta k}{\bar{k}}; \quad -2 < \varsigma < 2. \tag{26}$$

Only one of the field amplitudes is independent due to the relations (23a), (24a). Thus, any of them can be taken as the order parameter. For certainty, hereafter $\sigma_l(x)$ is considered as the ordering field.

It is worth noting that at $\varsigma \neq 0$ $N_f(x)$ and $c_l(x)$ depend non-linearly on $\sigma_l(x)$. For this reason $G(P,T)$ is not symmetric at the transformation $\sigma_l(x) \leftrightarrow \sigma_g(x)$, $\mu_l(P,T) \leftrightarrow \mu_g(P,T)$. Generally, the asymmetry parameter $\varsigma$ is not small.

The field $\tilde{\alpha}(x) = \alpha(x) - \alpha$ describes the fluctuations of the order parameter. Substituting the mean value of $\alpha(x)$ in (22), one obtains the mean chemical potential of the molecule in the heterophase state,

$$\mu(P,T) = \frac{\partial G(P,T)}{\partial N} = \bar{\mu}(P,T) + \frac{2+\varsigma}{4}\frac{1+2\alpha}{1+\varsigma\alpha}\Delta\mu(P,T) +$$

$$\frac{1}{\bar{k}(1+\varsigma\alpha)}\left\{-g_2\alpha^2 + \frac{1}{2}T\left[(1+2\alpha)\ln(1+2\alpha)+(1-2\alpha)\ln(1-2\alpha)-2\ln 2\right]\right\}. \tag{27}$$

Here

$$\bar{\mu}(P,T) = \frac{\mu_l(P,T)+\mu_g(P,T)}{2}; \quad \Delta\mu(P,T) = \mu_g(P,T)-\mu_l(P,T). \tag{28}$$

At $\varsigma = 0$ the expression (24) is isomorphic to the Hamiltonian of the non-symmetric Ising model with the external field $h(P,T) = -\Delta\mu(P,T)$. According to the definition given in Ref. 8 $\mu(P,T)$ is the effective Hamiltonian of the system.

A stable equilibrium state, corresponding to the minimum of the free energy, is determined by the equation of state

$$\frac{\partial \mu(P,T)}{\partial \sigma_l} = 0 \quad \text{with} \quad \frac{\partial^2 \mu(P,T)}{\partial \sigma^2} > 0. \tag{29}$$

Since $|\alpha| \leq 0.5$, in the vicinity of the critical point, a polynomial expansion of Eq. (27) in terms of $\alpha$ can be used. As result, the equation of state (29) reads (see Appendix A),

$$4\alpha\tau - 6\varsigma\tau\alpha^2 + \frac{16}{3}\alpha^3 = \frac{1}{T_c}\bar{k}h(P,T;\varsigma)(1+\varsigma\alpha)^2 + O(\tau_e\alpha^3). \tag{30}$$

Here

$$\tau = \frac{T-T_c}{T_c}; \quad T_c = g_2/2 \tag{31}$$

And

$$h(P,T;\varsigma) = -\frac{\partial \mu(P,T)}{\partial \alpha}\bigg|_{\alpha=0} = -\frac{4-\varsigma^2}{4T_c}\left[\mu_g(P,T)-\mu_l(P,T)\right] - \frac{T\varsigma\ln 2}{2\bar{k}}. \tag{32}$$

The solutions of Eq.(30) are well studied and can be found in books. Some of them are presented in Appendix A. Eq. (30) has three solutions. At $h(P,T;\varsigma) = 0$ below $T_c$ two of them,

$$\alpha_{1,2}(\tau) = \frac{9}{16}\varsigma\tau \pm \frac{1}{2}\sqrt{-\frac{3}{4}\tau} + O(\tau^{3/2}), \tag{33}$$

are stable. They describe the liquid-like and gas-like equilibrium states. The 3rd solution, $\alpha_3(\tau) = 0$, is unstable at $T < T_c$, but it is the only stable real solution at $T > T_c$.

As an example, the solutions (33) and the solution $\alpha_3(\tau) = 0$ are presented in Fig. 2 at $\varsigma = -1$. The $\alpha$-diameter,

$$\alpha_d(\tau) = (\alpha_1(\tau) + \alpha_2(\tau))/2 = \frac{9}{16}\varsigma\tau \tag{34}$$

is shown too. Above $T_c$ the line of thermodynamic equilibrium of the *l*- and *g*-fluctuons coincides with the solution $\alpha_3(P,T)$.

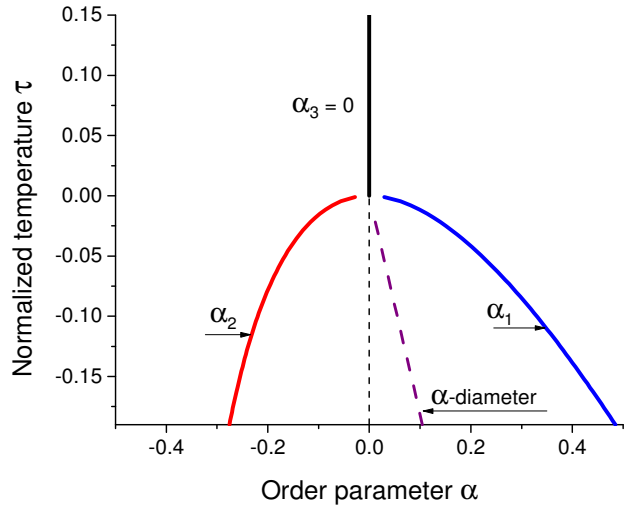

Fig. 2. The solutions of Eq. (30) vs normalized temperature $\tau$ at $h(P,T;\varsigma) = 0$ and $\varsigma = -1$. At $\tau < 0$ the stable solutions $\alpha_1$ and $\alpha_2$ are the branches of the phase coexistence curve. The $\alpha$-diameter $(\alpha_1(\tau) + \alpha_2(\tau))/2$) (thick dashed line) does not coincide with the symmetric solution $\alpha_3$ (thin dashed line). At $\tau > 0$ the line of the thermodynamic equilibrium of *l*- and *g*-fluctuons coincides with the solution $\alpha_3$.

## IV. MESOSCOPIC OZ EQUATION AND CORRELATIONS

Let us introduce the pair correlation function of the order parameter fluctuations,

$$\Lambda(\vec{r}) = \langle \tilde{\alpha}(\vec{x})\tilde{\alpha}(\vec{x}')\rangle = n_f \delta(\vec{r}) + n_f^2 \Gamma_f(\vec{r}); \quad \vec{r} = |\vec{x} - \vec{x}'|; \quad r > r_f \ . \tag{35}$$

Here $\Gamma_f(\vec{r})$ is the total correlation function of the fluctuons; $r_f$ is the mean distance between the fluctuons; $n_f$ is the density of fluctuons which is equal to

$$n_f = n_{f,l} + n_{f,g} = (N_{f,l} + N_{f,g})/V = N_f/V \tag{36}$$

The mesoscopic OZ equation, determining the fluctuonic direct correlation function, $\gamma_f(\vec{r})$, in analogy with Eq. (1), is following,

$$\Gamma_f(\vec{r}) = \gamma_f(\vec{r}) + n_f \int \gamma_f(\vec{r}')\Gamma_f(\vec{r}-\vec{r}')d^3\vec{r}' \tag{37}$$

where

$$\gamma_f(\vec{r}) = c_l \gamma_{f,l}(\vec{r}) + c_g \gamma_{f,g}(\vec{r}) = \frac{\gamma_{f,l}(\vec{r}) + \gamma_{f,g}(\vec{r})}{2} + \frac{[\gamma_{f,l}(\vec{r}) - \gamma_{f,g}(\vec{r})](\varsigma + 4\alpha)}{4(1+\varsigma\alpha)}. \tag{38}$$

The correlation function $\gamma_f(r)$ determines the fluctuonic short-range order. The fluctuonic OZ radius is

$$R_f^2 \sim \int r^2 \gamma_f(r)d\vec{r} \sim (r_{0,l} + r_{0,g})^2 \tag{39}$$

$\gamma_f(r)$ and $\Gamma_f(\vec{r})$ obey the OZ equations (2)-(4), but with the conformable replacement of $\rho$ and $R$ by the mesoscopic parameters $n_f$ and $R_f$, correspondingly. Hence,

$$1 + n_f \hat{\Gamma}_f(q) = \frac{1}{1 - n_f \hat{\gamma}_f(q)} \tag{40}$$

and

$$\hat{\gamma}_f(q) = n_f^{-1} - R_f^2 q^2 ; \quad q \to 0 \tag{41}$$

Consequently,

$$\hat{\Gamma}_f(q) = \frac{\xi_f^2}{1 + q^2 \xi_f^2} \tag{42}$$

and

$$\xi_f^2 = R_f^2 |\tau|^{-1} \text{ at } qR_f \ll 1 . \tag{43}$$

Accordingly,

$$\Gamma_f(\vec{r}) \sim r^{-1} \exp(-r/\xi_f). \tag{44}$$

It is worth noting that $\xi_f$ is scaled by $R_f$. The fluctuation theory of the criticality based on the effective Hamiltonian (27) leads to the correlation function of the universal form[9-14],

$$\Lambda_f(r) \approx \Gamma_f(r) \sim r^{-(1+\eta)} \exp(-\kappa r); \quad \kappa \sim R_f^{-1} |\tilde{\tau}|^\nu . \tag{45}$$

Here the critical exponent $\nu$ is nearly equal to 2/3. The parameter $\eta$ in the tree-dimensional system is small, $\eta \approx 0.06$ [11]. The critical temperature in (45) differs from its classic value,

$$T_c \to \tilde{T}_c = T_c - \Delta T_c ; \qquad \tau \to \tilde{\tau} = \frac{T - \tilde{T}_c}{\tilde{T}_c} \tag{46}$$

The shift of the critical temperature $\Delta T_c$ appears due to renormalization. It is comparable with the temperature range of the fluctuation region.

The Fourier transform of $\Gamma_f(\vec{r})$ is as follows,

$$\widehat{\Gamma}_f(q) \sim \frac{\kappa^{\eta-2}}{\left(1+q^2/\kappa^2\right)^{\frac{1-\eta}{2}}} \tag{47}$$

Let us use Eqs. (40) and (47) to calculate $\widehat{\Gamma}_f(q)$, $\widehat{\gamma}_f(q)$ and their correlation lengths, $\xi_f$ and $\tilde{R}_f$, at $q/\kappa \ll 1$, i.e. within the fluctuation range where the length $\tilde{R}_f$ is not equal to $R_f$ determined by Eq. (39). The calculations give the following result,

$$\widehat{\Gamma}_f(q) \sim \kappa^{\eta-2}\left[1-q^2\kappa^{\eta-2}(1-\eta)/2\right], \quad q/\kappa \ll 1, \tag{48}$$

and

$$\xi_f^2 = -\frac{1}{2\widehat{\Gamma}(0)}\frac{\partial^2\widehat{\Gamma}(q)}{\partial q^2}\bigg|_{q=0} = \frac{1-\eta}{2\kappa^2}, \quad \tilde{R}_f^2 = -\frac{1}{2\widehat{\gamma}_f(0)}\frac{\partial^2\widehat{\gamma}_f(q)}{\partial q^2}\bigg|_{q=0} \sim \left(\xi_f\right)^\eta \tag{49}$$

Taking into account that $\kappa \sim R_f^{-1}|\tilde{\tau}|^\nu$, we have

$$\tilde{R}_f \sim R_f |\tilde{\tau}_e|^{-\frac{\nu\eta}{2}}. \tag{50}$$

One can see that $\tilde{R}_f$ is infinite at $\tilde{\tau}=0$. Hence, the OZ hypothesis fails in the vicinity of the critical temperature. Fisher obtained this result combining the correlation function of the density fluctuations in the form (45) and the OZ equation (1).[11]

It has to be noted that due to smallness of the exponent in (50), $\nu\eta \approx 0.04$, $\tilde{R}_f$ decreases fast with the $|\tilde{\tau}_e|$ increase and becomes nearly equal to $R_f$ out of a narrow temperature range. The experimental data show that the classic 3D models are acceptable at $|\tilde{\tau}| > |\tilde{\tau}_0| > 0.1$.[14,30] Determining the fluctuation range as the region where the $\tilde{R}_f$ deviates considerably from $R_f$, we can estimate the width of this range. As it follows from Eq. (50)

$$\frac{\tilde{R}_f}{R_f} \sim \left|\frac{\tilde{\tau}_e}{\tilde{\tau}_0}\right|^{-\frac{\nu\eta}{2}}. \tag{51}$$

The graphic representation of this ratio at $|\tilde{\tau}_0|=0.1$ is shown in Fig. 3. One can see that $\tilde{R}_f$ undergoes a noticeable change at $|\tilde{\tau}|<0.025$. At $|\tilde{\tau}|=0.025$ we have $\tilde{R}_f(0.025) \approx 0.96 R_f$, i.e. the deviation of the relative value of $\tilde{R}_f$ from $R_f$ becomes comparable with $\tilde{\tau}$ at $|\tilde{\tau}|=0.025$ and increases with $\tilde{\tau}$ decrease. The significant deviation of $\tilde{R}_f$ from $R_f$ occurs at $\tilde{\tau} < 0.01$. Accordingly, the shift of the critical temperature $\Delta T_c$ in Eq. (46) is $\sim 10^{-2} T_c$.

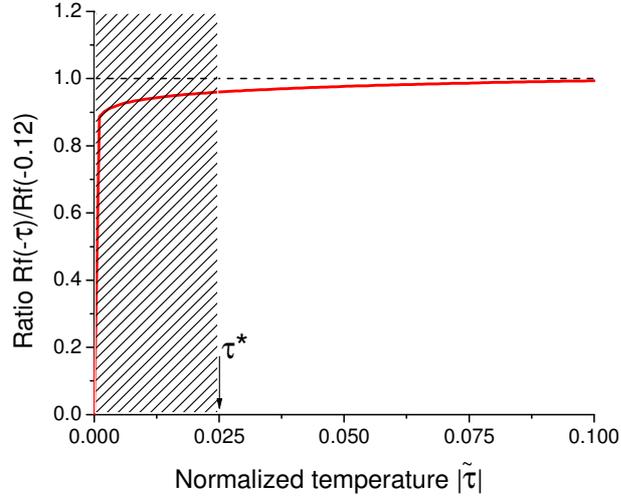

Fig. 3. The correlation length of the direct correlation function vs $|\tilde{\tau}|$. At $|\tilde{\tau}| = 0.025$ the deviation of $\tilde{R}_f$ from its value at $|\tilde{\tau}| = 0.12$ is equal to 5%. The fluctuation region $|\tilde{\tau}| < 0.025$ is shaded.

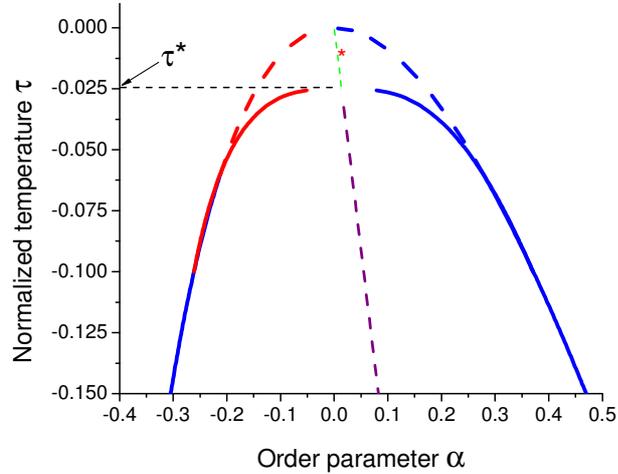

Fig. 4. The phase coexistence curve modified by the critical fluctuations. The classic coexistence curve near $T_c$ is shown by the dashed line. The critical temperature shift, $\tau^* = \Delta T_c / T_c$, is indicated.

The phase coexistence curve modified by the critical fluctuations is shown in Fig. 4. It is taken into account that the leading terms of the renormalized solutions $\alpha_{1,2}(\tau)$ are proportional to $\pm(-\tilde{\tau})^{1/3}$. The classic and modified parts of the curve and their first derivatives are joined at $\tilde{\tau} \approx 0.06$.

## V. ρ-DIAMETER

The experimentally measured gas-liquid coexistence curve is usually presented in the specific volume-temperature or in the density-temperature variables since the density plays the role of the order parameter in known models of the gas-liquid critical point. To get the equation for the coexistence curve on the $(\rho,\tau)$-plane we have to use the relation of $\rho$ with the order parameter $\alpha$ (B.1)-(B.5) (Appendix B)

$$\rho = \frac{4-\varsigma_v^2}{4-\varsigma\varsigma_v}\left[1+D_1\alpha+D_2\alpha^2\right]\bar{\rho}+O(\alpha^3). \tag{52}$$

$$D_1 = \frac{\varsigma_v(4-\varsigma^2)}{4-\varsigma\varsigma_v}; \qquad D_2 = D_1^2 - \varsigma D_1; \tag{52a}$$

$$\varsigma_v = \frac{\rho_l-\rho_g}{\bar{\rho}} = \frac{v_g-v_l}{\bar{v}}; \quad 0<\varsigma_v<2; \quad \bar{\rho} = \frac{\rho_l+\rho_g}{2} \tag{52b}$$

The branches of the coexistence curve on the $(\rho,\tau)$-plane are related to the branches of the α-coexistence curve $\alpha_1$ and $\alpha_2$,

$$\rho_{1,2}(\tau) = \frac{4-\varsigma_v^2}{4-\varsigma\varsigma_v}\left[1+D_1\alpha_{1,2}(\tau)+D_2\alpha_{1,2}^2(\tau)\right]\bar{\rho}(\tau)+O(\alpha_{1,2}^3) \tag{53}$$

The classic values of $\alpha_1$ and $\alpha_2$ are determined by Eq. (33). Hence, the ρ-diameter is determined by the following equation,

$$\rho_d(\tilde{\tau}) = \frac{1}{2}\left[\rho(\alpha_1(\tilde{\tau}))+\rho(\alpha_2(\tilde{\tau}))\right] = \left[1+\left(\frac{\bar{\rho}'}{\bar{\rho}(0)}+\frac{3}{16}D_1\varsigma-\frac{3}{4}D_2\right)\tilde{\tau}\right]\rho_c \tag{54}$$

$$\rho_c = \frac{4-\varsigma_v^2}{4-\varsigma\varsigma_v}\bar{\rho}(T_c,P_c); \qquad \bar{\rho}' = \frac{\partial\bar{\rho}(\tilde{\tau})}{\partial\tilde{\tau}} \tag{55}$$

The relation (54) reproduces the so-called Cailletet-Mathias empirical rectilinear law[30] in the classic limit. Within the fluctuation region this equation fails. In this case the polynomial representation of the ρ-diameter is as follows,[8,12,14-19]

$$\frac{\rho_d(\tilde{\tau})-\rho_c}{\rho_c} = \left(\frac{\bar{\rho}'}{\bar{\rho}(0)}+\frac{3}{16}\varsigma D_1\right)\tilde{\tau}+D_{2\beta}\tilde{\tau}^{2\beta}+D_{1-\alpha}\tilde{\tau}^{1-\alpha}+.... \tag{56}$$

Here $\alpha \approx 0.1$ is the critical exponent which determines the singularity of the heat capacity at $T \to T_c$.

The coefficients $D_{2\beta}$ and $D_{1-\alpha}$ appear due to the violation of the Hamiltonian symmetry. They can be calculated using the scaling analysis.[8,14-18] To use it, the Hamiltonian (A.1) has to be rewritten in terms of the order parameter

$$\eta(P,T) = \frac{\rho(P,T) - \rho_c}{\rho_c} \tag{57}$$

Eq. (B.12) with coefficients (B.18) represents the desired result. Then the coefficients $D_{2\beta}$ and $D_{1-\alpha}$ are connected by the expressions (B.13), (B.14) with the coefficients of the Hamiltonian (B.12). Finally we have (see (B.19), (B.21)).

$$D_{2\beta} = -\frac{3}{4} D_1 \varsigma, \tag{58}$$

$$D_{1-\alpha} = \frac{3}{4}(D_2 + D_1\varsigma) + 0.95\varsigma \tag{59}$$

For the Yang-Yang ratio[31], $Y_R$, we have (see (B.21)),

$$Y_R = -\frac{\varsigma}{D_1} \tag{60}$$

$$Y_R = -\frac{\varsigma}{D_1} \tag{60}$$

## VI. THERMODYNAMIC MANIFESTATIONS OF THE HPF EFFECTS

Let us consider the effects of the HPF on the thermodynamic quantities within the critical and supercritical region. By differentiating the free energy one can obtain the following expressions for the specific volume, entropy and their derivatives.

The specific volume:

$$v = c_l v_l + (1 - c_l) v_g \tag{61}$$

The specific entropy:

$$s = c_l s_l + (1 - c_l) s_g - \frac{1}{\bar{k}(1+\varsigma\alpha)} \left( \sigma_l \ln \sigma_l + \sigma_g \ln \sigma_g \right) = c_l s_l + (1 - c_l) s_g + O(\bar{k}^{-1}) \tag{62}$$

The thermal expansion coefficient at constant $P$:

$$\alpha_P = \frac{\partial \ln v}{\partial T} = \frac{1}{v}\left(c_l v_l \alpha_{P,l} + c_g v_g \alpha_{P,g}\right) + \frac{v_l - v_g}{v}\frac{\partial c_l}{\partial \alpha}\frac{\partial \alpha}{\partial T}\bigg|_P \tag{63}$$

The compressibility at constant T:

$$\kappa_T = \frac{\partial \ln v}{\partial P} = \frac{1}{v}\left(c_l v_l \kappa_{T,l} + c_g v_g \kappa_{T,g}\right) + \frac{v_l - v_g}{v}\frac{\partial c_l}{\partial \alpha}\frac{\partial \alpha}{\partial P}\bigg|_T \tag{64}$$

The heat capacity at constant pressure:

$$C_p = T\frac{\partial s}{\partial T} = c_l C_{pl} + (1-c_l)C_{pg} + \left(s_l - s_g\right)\frac{\partial c_l}{\partial \alpha}\frac{\partial \alpha}{\partial T} \tag{65}$$

The heat capacity at constant volume:

$$C_v = C_p + T(v\alpha_P)^2/(v\kappa_T): \tag{66}$$

Here $\kappa_{T,l}$, $\kappa_{T,g}$ is the compressibility of a liquid and a gas at constant temperature; $\alpha_{P,l}$, $\alpha_{P,g}$ are the thermal expansion coefficients of a liquid and a gas at constant pressure;

$$\frac{\partial c_l}{\partial \alpha} = \frac{4-\varsigma^2}{4}\frac{1}{(1+\varsigma\alpha)^2}; \tag{67}$$

$$\frac{\partial \alpha}{\partial T}\bigg|_P = \bar{k}\frac{(4-\varsigma^2)(s_l - s_g)}{16T_c(-\tau + 4\alpha^2)}; \tag{68}$$

$$\frac{\partial \alpha}{\partial P}\bigg|_T = \bar{k}\frac{(4-\varsigma^2)(v_l - v_g)}{16T_c(-\tau + 4\alpha^2)}. \tag{69}$$

The last terms of the right-hand sides of the relations (63)-(66) are singular at $\tau \to 0$, $\alpha \to 0$, as it is seen from the formulae (67)-(69) which reproduce the results of the classic theory of the critical point but taking into account the asymmetry of the HPF model. Within the fluctuation range these quantities obey the universal criticality laws.[9-14] The non-singular terms in (63)-(66), which describe the effects of the short-range heterogeneities, are nonuniversal. These effects are dominating beyond the fluctuation range.

## VII. STRUCTURE FACTOR

The correlation function of the critical long-range density fluctuations, $\Gamma(q)$, as it follows from the fluctuation theory of critical point applied to the effective Hamiltonian (B.12), (B.18) formulated in terms of the order parameter $\eta(P,T)$ (57), has the standard form (47),

$$\hat{\Gamma}(q) \sim \frac{\kappa^{\eta-2}}{\left(1+q^2/\kappa^2\right)^{\frac{1-\eta}{2}}}; \quad qR_f \ll 1 \quad . \tag{70}$$

Considering the heterophase state as a fluctuonic mixture with the fractions of *l*- and *g*-fluctuons equal to $\sigma_l$ and $1-\sigma_l$ respectively, we can represent the mesoscopic static structure factor as follows,

$$S(q) = n_f \left[ \sigma_l |u_l(q)|^2 + (1-\sigma_l)|u_g(q)|^2 \right]; \quad qR_f > 1. \tag{71}$$

Here $|u_l(q)|^2$ and $|u_g(q)|^2$ are the form-factors of *l*- and *g*-fluctuons which are determined thus,

$$|u_l(q)|^2 = k_l |a(q)|^2 \left[ 1 + \rho_l \hat{\gamma}_{f,l}(q) \right]; \quad qr_{0l} > 1; \tag{71a}$$

$$|u_g(q)|^2 = k_g |a(q)|^2 \left[ 1 + \rho_g \hat{\gamma}_{f,g}(q) \right]; \quad qr_{0g} > 1 \tag{71b}$$

In these relations $|a(q)|^2$ is the form-factor of the molecule. $\hat{\gamma}_{f,l}(q)$ and $\hat{\gamma}_{f,g}(q)$ are the Fourier transforms of the short-range parts of the density-density correlation functions of the *l*- and *g*-fluctuons respectively. Since the size of the fluctuon is comparable with the correlation length of the direct correlation function, the $\hat{\gamma}_{f,l}(q)$ and $\hat{\gamma}_{f,g}(q)$ are nearly equal to the direct correlation functions of the states with densities equal to $\rho_l$ and $\rho_g$, correspondingly.

Combining equations (4), (52), (70), (71) one gets the equations determining the critical structure factor taking into account its short-range part and long-range limits

$$S(q) \approx n_f \left[ \sigma_l |u_l(q)|^2 + (1-\sigma_l)|u_g(q)|^2 \right] + \left\{ \frac{4-\varsigma_v^2}{4-\varsigma\varsigma_v} \left[ 1 + D_1 \alpha + D_2 \alpha^2 \right] \bar{\rho} \right\}^2 \Gamma(q) \tag{72}$$

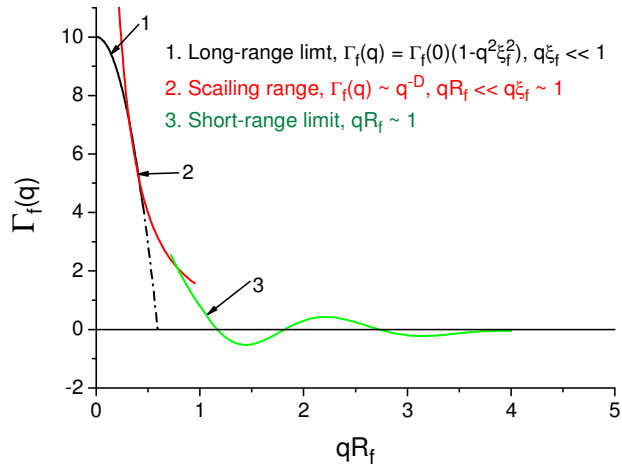

Fig.5. A schematic representation of the fluctuon-fluctuon correlation function vs $qR_f$ comprised of the functions (72)-(74).

The represented in Fig.5 total correlation function of the heterophase fluid includes both the universal critical (long-wave) part (70) and the nonuniversal short-wave summand (71). At $q\xi_f \ll 1$

$$\hat{\Gamma}(q) \approx \left(\frac{\xi_f}{R_f}\right)^{2-\eta} \left(1 - q^2 \xi_f^2\right). \tag{73}$$

and within the scaling range, at $qR_f << q\xi_f \sim 1$,

## VIII. DISCUSSION

The HPF model, based on the phenomenogical effective Hamiltonian (27) which follows from the free energy expression (22)-(24), is formulated in terms of the interacting mesoscopic entities – liquid-like and gas-like fluctuons which are identified as the smallest transient species of the matter possessing characteristic properties of the liquid and gaseous phase respectively. As it was pointed out, the heterophase structure of the matter in the vicinity of the gas-liquid critical point was for the first time conjectured by van der Waals.[2] Later this idea was reincarnated scores of times in different forms.[6,7, 32-40]

Bernal, who constructed the dense random packing model of the amorphous matter, emphasized that the liquid-gas coexistence curve, continued in the overcritical region (Bernal called it *the hypercritical line*) delimits the liquid-like and gas-like states. The associativity of the gaseous phase increases continuously with the density increase until the hypercritical line is passed and the closely packed liquid structure is formed. The concentration of the molecular microaggregates grows with the mean density increase but the "holes" do not disappear even within the solidified glassy state.

On a qualitative level, the HPFM is in harmony with Bernal's idea but the *g*-fluctuons (meso-bubbles) are the gas-filled holes. The line $\alpha(P,T) = 0$ at $T > T_c$ (see Fig.2) can be identified as the hypercritical one. On this line *l*- and *g*-fluctuons are in the thermodynamic equilibrium and their numbers, $N_{f,l}$ and $N_{f,g}$, are equal. At temperature decrease or pressure increase $N_{f,l}$ is growing. The growth rate is proportional to $\tau^{-1}$, as the solution (A.13) shows.

Considering supercritical fluids, Semenchenko educed that a heterophase gas-liquid mesophase is formed during a continuous isothermal transformation of the gaseous state into the liquid one under pressure.[34,35] He noted that the Frenkel droplet model can not be used to describe the mesophase. [35] Semenchenko assumed that the mesophase domain on the phase diagram is bounded by a "quasi-spinodal" which is determined by the loci of the inflection points on the isotherms. The "quasi-spinodal" has a parabolic shape on the (*P,V*)-plane. The high-pressure branch of this line is close to the gas-liquid phase coexistence curve continued into the supercritical region.

As it follows from the van der Waals equation, the "quasi-spinodal" in terms of the scaled *P,V* is determined by the following equation,

$$\tilde{P} = \frac{1}{\tilde{V}^2}\left[\left(\frac{3\tilde{V}-1}{\tilde{V}}\right)^2 - 3\right]; \quad \tilde{P} = \frac{P}{P_c}; \quad \tilde{V} = \frac{V}{V_c} \quad (75)$$

According to the law of corresponding states, this equation has to find the loci of the inflection points of fluids as accurate as the normalized van der Waals equation describes their other properties. Since the "quasi-spinodal" line can be found experimentally, one can check the universality of the Eq. (75). It occurs that the experimental data are in poor harmony with this equation.[35] The gas-liquid mesophase domain is considered below in the framework of the HPFM.

The data on the slow neutrons scattering on the supercritical water were recently interpreted in terms of a two-state model based on Semenchenko's ideas.[37]

The Bernal's ideas constitute the basis of the research done by Woodcock[38,39] and Finney and Woodcock[40]. Considering the heterophase bubble-droplet states, they separated out the three types of these states with different topologies. In the liquid-like state the bubbles exist as the isolated (as it is shown in Fig.2a) or aggregated g-fluctuons but the volume fraction of the bubbles is not enough to form the spanned (percolating) domain. The topology of the gas-like heterophase state is a "mirror reflection" of the liquid-like state topology: the droplets can form the domains of different finite sizes but the total volume fraction of the droplets is less than the percolation threshold. The threshold volume fraction for percolation in the 3D binary liquid mixture is found to be nearly equal to 0.16.[41,42] Thus one can conclude that the volume fraction of the droplets is less than 0.16 in the gas-like states while in the liquid-like states it exceeds 0.84.

In the state of the third type, which is called "mesophase", the liquid-like and gas-like fractions form the interpercolating spanned domains. On the phase diagrams the mesophase is separated by the droplet percolation line (DPL) and the bubble percolation line (BPL) from the gas-like and the liquid-like phase, correspondingly.

It is assumed that the thermodynamic potentials are not analytical and that weak continuous phase transitions take place on both percolation lines (the type of these transitions is yet unknown). Thus, it is implied that the DPL and BPL are the boundaries of phases specified by the topology of the HPF. A procedure of determining the loci of DPL and BPL on phase diagrams was proposed by Woodcock.[38,39] It is based on the analysis of the experimental data and the results of computer simulation. In particular, the mesophase domain is filled by the nearly rectilinear segments of the isotherms above $T_c$.

Since the HPFM describes the heterophase states with arbitrary amounts of the gas-like and liquid-like fractions, it allows to get the equations for the DPL and BPL. The volume fraction occupied by the droplets is proportional to $N_{f,l}$ which is equal to $(0.5+\alpha)N_f$. Therefore, the percolation thresholds are connected with specific values of the order parameter $\alpha$. As it follows from (B.1), the liquid-like volume fraction, $\varpi_l$, is equal to

$$\varpi_l(P,T) = \frac{c_l v_l(P,T)}{v(p,T)} = \frac{(2+\varsigma)(2-\varsigma_v)}{2} \frac{1+2\alpha(P,T)}{4-\varsigma_v\varsigma+4(\varsigma-\varsigma_v)\alpha(P,T)} \qquad (76)$$

Hence, as it was noted, at

$$\varpi_l(P,T) > \varpi_{l,d} = 0.16 \qquad (77)$$

the droplets form the percolation domain. At

$$\varpi_l(P,T) < \varpi_{l,b} = 0.84 \qquad (78)$$

the bubbles are percolating. The "mesophase" exists within the range $0.16 < \varpi_l(P,T) < 0.84$.

Substitution of $\varpi_{l,d}$ and $\varpi_{l,b}$ in Eq. (76) allows to find the values of the order parameter at the percolation thresholds, $\alpha_d$ and $\alpha_b$,

$$\alpha_{d(b)} = \left[2+2\frac{4(\varsigma_v-\varsigma)\eta_{l,d(b)}}{(2+\varsigma)(2-\varsigma_v)}\right]^{-1}\left[\frac{(4-\varsigma_v\varsigma)}{(2+\varsigma)(2-\varsigma_v)}\eta_{l,d(b)}-1\right] \qquad (79)$$

As an example, in Fig. 6 the volume fraction $\varpi_l$ vs order parameter $\alpha$ at $\varsigma = 0.2$ and $\varsigma_v = 0.7$ is represented. The values of the asymmetry parameters are chosen taking into account the bounds of their ranges considered in Appendix C. Indicated in Fig. 6 values of the order parameter at the percolation thresholds are: $\alpha_d \approx -0.25$ and $\alpha_b \approx 0.38$.

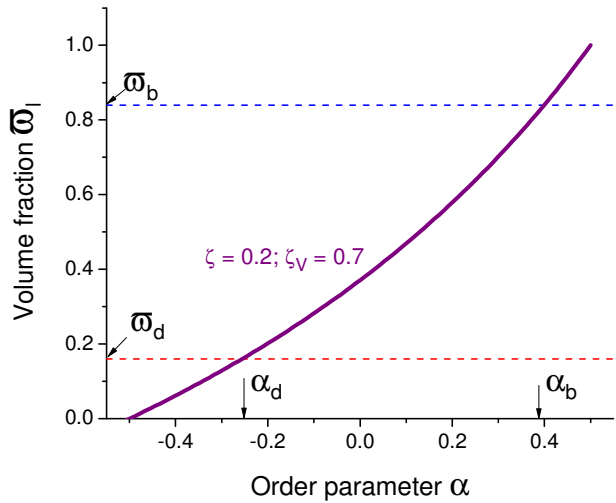

Fig. 6. The volume fraction of the droplets vs the order parameter $\alpha$ at $\varsigma = 0.25$ and $\varsigma_v = 1$. The percolation thresholds $\varpi_{l,d}$ and $\varpi_{l,b}$ are indicated. The values of the order parameter at the percolation thresholds, $\alpha_d$ and $\alpha_b$, are indicated by the arrows too.

By substituting the threshold values $\alpha_d$ and $\alpha_b$ into the equation of state (30) one obtains the equations determining the DPL and BPL loci on the $(P,T)$-plane in the applicability region of the classical theory.

To get the loci of the percolation lines on the $(\rho,T)$-plane, $\alpha_d$ and $\alpha_b$ have to be substituted into Eq. (52) considering $\bar{\rho}(T)$ as the phenomenological function which has to be taken from experimental data. It occurs that the results of experimental measurements of the normalized gas and liquid density,

$$\eta_g(\tau) = \frac{\rho_g(\tau) - \rho_c}{\rho_c} \quad \text{and} \quad \eta_l(\tau) = \frac{\rho_l(\tau) - \rho_c}{\rho_c} \tag{80}$$

below $T_c$ for many simple liquids can be properly fitted by the master curve which is a cubic parabola,[43]

$$\eta_g(\tau) = 1.8\tau^{1/3}; \quad \eta_l(\tau) = -1.8\tau^{1/3} \tag{81}$$

The diameter of the parabola is rectilinear,

$$\eta_{\text{diameter}}(\tau) = \rho_c(1 - 0.75\tau) \tag{82}$$

Let us note that Eq.(81) gives the correct critical exponent.

We can use the empiric formulas (81) to find the cross points of DPL and BPL with $\eta_g(\tau)$ and $\eta_l(\tau)$ at $\tau < 0$ (Appendix D). The cross points are the end points of DPL and BPL because the stable threshold points do not exist within the region of phase metastability.

The coordinates of the end points are correspondingly equal to

$$\eta_d = D_1\alpha_d + D_2\alpha_d^2 < 0; \quad \tau_d = \left(\frac{D_1\alpha_d + D_2\alpha_d^2}{1.8}\right)^3 \sim -10^{-3} \tag{83}$$

and

$$\eta_b = D_1\alpha_b + D_2\alpha_b^2 > 0; \quad \tau_b = -\left(\frac{D_1\alpha_b + D_2\alpha_b^2}{1.8}\right)^3 \approx -10^{-2} \tag{84}$$

The DPL and BPL fitted to the experimental data[43] at $\varsigma = 0.2$ and $\varsigma_v = 0.7$ are represented in Fig. 7. The line $\bar{\rho}(\tau)$ is shown too. Its slope is determined using Eq.(81) (see Appendix C).

The loci of DPL and BPL on the $(\pi, \tau)$-plane ($\pi = (P - P_c)/P_c$ is the normalized pressure) is represented in Fig. 8 at the same values of parameters $\varsigma$ and $\varsigma_v$. The slope of the coexistence curve, $\pi'_e$, is taken to be equal to 4.

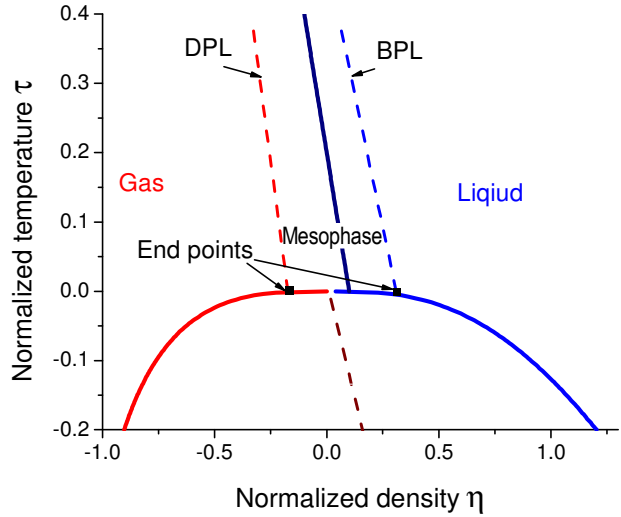

Fig. 7. Loci of the DPL and BPL on the $(\rho,\tau)$-phase diagram. The ρ-diameter at $\tau<0$ is represented by the dashed line. The line $\bar{\rho}(\tau)/\bar{\rho}(0)$ is shown by the solid line at $\tau>0$. The cross points (82), (83), which are the end points of the percolation lines, are indicated by arrows.

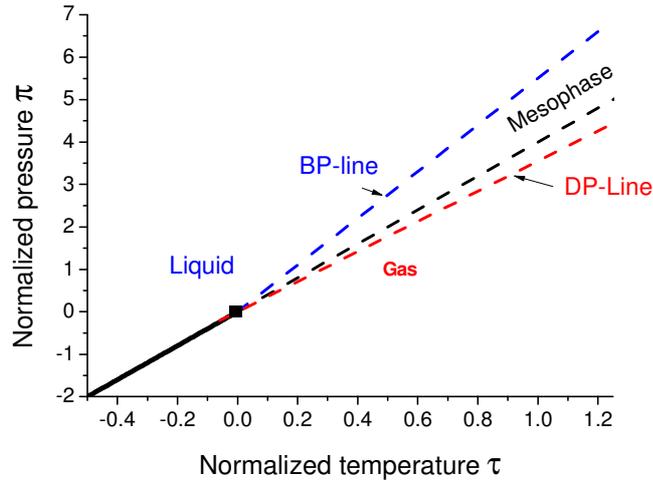

Fig. 8. The loci of DPL and BPL on the $(\pi,\tau)$-plane at $\varsigma=0.2$ and $\varsigma_v=0.7$, and $\pi'_e=4$.

The droplets in a gas near the DPL, before the percolating domain is formed, are the fractals with the fractal dimension $D_f \approx 2.5$.[44] Their size distribution is similar to the Fisher droplet distribution (10). It has the same exponent but the number of molecules is replaced by the number of fluctuons per droplet, $k_f$,

$$f(k_f) \sim k_f^{-\tau} \tag{85}$$

The gas-like domains within the liquid at $\varpi_l(P,T)$ approaching the bubble percolation threshold, $\varpi_{l,b} = 0.84$, from above also have a similar topology and size distribution.

Bernal noted that the dynamics of a liquid and a gas are essentially different. Hence this difference can be used to identify the liquid-like and gas-like heterophase states. The most pronounced variation of the short-wave dynamics is expected on the scales comparable with $R_f$. The results of research reported in [45-49] confirm the conjecture that the dynamics of the fluid is very sensitive to the density/pressure changes. This feature allows to discriminate the dynamics of the liqud-like and gas-like states.

The dynamic structure factor of the supercritical Ar was investigated by means of the inelastic X-ray scattering.[45] As a result of the research, the positive sound dispersion (PSD) is revealed and attributed to the liquid-like species. The PSD is a viscoelastic effect which consists in a positive shift of the acoustic collective excitation frequency as compared to that prescripted by the hydrodynamic dispersion law. The PSD reduces with the decrease of the pressure/density of the fluid.

The classical MD simulations of the Ar dynamics above $T_c$ also show that the PSD is significant at high pressure/density. At low pressure it vanishes completely. The crossover range is located near the Widom line. Since the PSD is a quantity which is considerable for the liquid-like states and is small for the gas-like fluids, it can be used as the indicator of the fluid associativity. The analytical model of the short-range fluid dynamics was developed by Bryk et al.[46] Further progress in this field was achieved.[47-49] The hypercritical line is determined in a wide temperature range, up to $T/T_c \approx 5$.

It has to be noted that the renowned lattice model of Lee and Yang (LY) [50,51] reduces the problem of the gas-liquid critical point to the classic two-state Ising model. In the LY model the lattice sites can be vacant or occupied by the mutually gravitating atoms. It is in essence a symmetric model and the asymmetry effects are beyond its scope (as it was pointed out by the authors). Formally the equation of state of the HPFM (30) transforms into the LY model equation of state at $\varsigma = 0$ and the modified expression for the external field. Evidently, the analogy between the HPFM and LY model is rather limited in extent.

The majority of models devoted to the thermodynamics of phase transitions are formulated in the two-state (two-species) approximation.[5] The scope of this approximation caused by the multiplicity of the statistically significant species of different nature (multiplicity of the HPF types) can be expanded analogously to those performed in Refs. 26-28. Such generalization of the HPFM becomes necessary, e.g. while accounting for the fluids containing additions or colloidal inclusions.

The temperature range of the crossover from the fluctuation to classical behavior of the total correlation function determined in Sect.4 supplements the Levanyuk-Ginzburg criterion of the classical theory applicability.[52,53] The Ginzburg temperature $T_G$ of the gas-liquid transition was estimated within different classical models.[54-58] The estimations show that $T_G \sim 10^{-2} - 10^{-1} T_c$ depending on the values of the model parameters. In particular, the temperature $T_G$ of ionized fluids decreases with the increase of the ionic component.

This result can be qualitatively explained as follows. Patashinskii and Pokrovsky introduced the interaction radius of the fluctuations of order parameter, $r_{interaction}$, which is a constant in terms of the coefficients of the Hamiltonian (27).[8,12] The Levanyuk-Ginzburg criterion, which follows from (27), reads

$$\tau \gg \tau_G \sim (r_{interaction} / R_f)^6; \quad \tau_G = \frac{T_c - T_G}{T_c} \tag{86}$$

This relation shows that $T_G$ decreases with the $R_f$ increase. Since the Debye-Hűckel radius, which is a characteristic correlation length in the ionized gas, can be much larger than the interatomic distance, one can expect a decrease of $T_G$ with the ionicity increase.

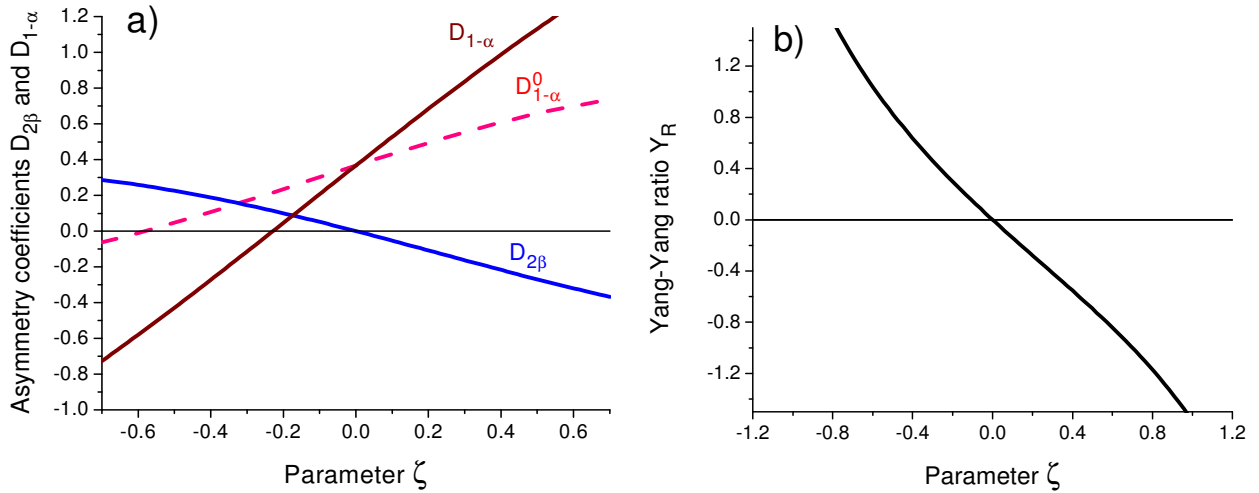

Fig. 9 The coefficients $D_{2\beta}$, $D_{1-\alpha}$ and $D_{1-\alpha}^0$ vs the parameter $\varsigma$ are represented in panel a). The coefficient $D_{1-\alpha}^0$ is calculated neglecting the contribution of external field nonlinearity (the term $\sim h\eta^2$). The Yang-Yang ratio $Y_R$ vs the parameter $\varsigma$ is shown in panel b).

Eq. (55) shows that the classic ρ-diameter is rectilinear but the asymmetry of the mesoscopic Hamiltonian (27) leads to the non-linearity of the diameter within the fluctuation range (Eq.(56)). As Eqs. (58)-(60) show, the nonuniversal coefficients $D_{2\beta}$, $D_{1-\alpha}$ and the Yang-Yang ratio, $Y_R$, associated with them are not equal to zero at $\varsigma, \varsigma_v \neq 0$. The graphic representation of these coefficients at $\varsigma_v = 0.7$ is shown in Fig. 9. In panel a) the coefficients $D_{2\beta}$ and $D_{1-\alpha}$ are represented. To underline the importance of the contribution of the non-linear field term, $h\eta^2$, in $D_{1-\alpha}$, the coefficient $D_{1-\alpha}^0$, which is calculated at $h\eta^2 = 0$, is shown too. As can be seen,

- $|D_{2\beta}| < 1$;
- the ratio $D_{2\beta}/D_{1-\alpha}$ is negative at $\varsigma > 0$ and at $\varsigma < -0.2$ and its absolute value varies within a wide range;
- the sign of $Y_R$ coincides with the sign of $\varsigma$ and its absolute value grows fast with the parameter $\varsigma$ increase.

One can expect that these results are reasonable for fluids with the $(\rho, \tau)$-phase diagram similar to that obtained in Ref. 43.

The results of accounting for the asymmetry of fluid criticality, which are summarized in the overview[18], are obtained in a "microscopic" approach which is based on the polynomial representation of the effective Hamiltonian with coefficients depending on the parameters of the molecular pair potential. The values of the coefficients $D_{2\beta}$, $D_{1-\alpha}$, and $Y_R$ obtained above belong to approximately the same ranges as those found by Weiss et al.[18]

The HPFM can be verified by comparing the deduced formulas with the experimental data. This issue is important for clarifying the nature of the supercritical fluid state. The revealed interplay of the macroscopic and mesoscopic thermodynamics and structure properties of a fluid serves as a ground for the HPFM validation.[d]

As it follows from (61), (62), in the mesophase domain on the $(P,T)$-plane

$$c_l(P,T) \approx \frac{v(P,T) - v_g(P,T)}{v_l(P,T) - v_g(P,T)} \tag{87}$$

and, on the other hand,

$$c_l(P,T) \approx \frac{s(P,T) - s_g(P,T)}{s_l(P,T) - s_g(P,T)}. \tag{88}$$

To use the formula (87) for finding the order parameter, $c_l(P,T)$, experimentally it is necessary to measure the specific volume $v(P,T)$ in a wide $(P,T)$-domain which also includes the mesophase domain. Beyond the mesophase domain $v(P,T)$ is nearly equal to $v_l(P,T)$ (within the domain of the liquid-like phase existence) or to $v_g(P,T)$. The extrapolation of these functions into the mesophase domain gives the values of $v_l(P,T)$ and $v_g(P,T)$ which can be substituted in Eq. (87). Analogous procedure can be applied when using formula (88).

---

[d] The validation procedure of the HPFM, similar to the one described below, was proposed and used to investigate the HPF effects in the supercooled liquids.[27, 59-61]

Let us note that the order parameter varies from ~0.1 to ~0.9 on crossing the mesophase domain while the specific volumes $v_l(P,T)$ and $v_g(P,T)$ presumably change much less.

Using the extrapolated quantities to determine the order parameter we ignore the proximity effects of the order of $a/r_{0g}$ which appear due to the mesoscopic size of the fluctuons. To estimate the volume proximity effect one can compare the measured value of the specific volume on the coexistence curve, $v_{exp}(P_e, T_e)$, with its the extrapolated value, $v_{ext}(P_e, T_e)$,

$$v_{ext}(P_e, T_e) = \frac{4 - \varsigma \varsigma_v}{4} \overline{v}_{ext}(P_e, T_e) \tag{89}$$

These relations can be used to validate the HPFM of the fluid. In particular, the validation of formulae (25), (A.14) which determine $c_l(P,T)$ can be performed.

A similar validation procedure can be also performed using the experimental measurements of the structure factor of supercritical fluid. The structure factor in absence of the long-range density-density correlations is determined by Eqs. (71-71b). Taking into account that $n_f k_l \sigma_l = c_l$ and $n_f k_g (1-\sigma_l) = c_g$, we see that $S(q)$ is the weighted sum of the structure factors of the liquid-like and gas-like fractions, $S_l(q)$ and $S_g(q)$.

$$S(q) = c_l S_l(q) + (1 - c_l) S_g(q) \tag{90}$$

$$S_l(q) = |a(q)|^2 \left[1 + \rho_l \hat{\gamma}_{f,l}(q)\right] \tag{90a}$$

$$S_g(q) = |a(q)|^2 \left[1 + \rho_g \hat{\gamma}_{f,g}(q)\right] \tag{90b}$$

Noting that $S(q;P,T) \approx S_g(q;P,T)$ at $c_l(P,T) \ll 1$ and $S(q;P,T) \approx S_l(q;P,T)$ at $1 - c_l(P,T) \ll 1$, we see that the structure factor measured beyond the mesophase domain, i.e. within the domains of the liquid and gaseous phases shown in Fig.7 and Fig.8, is nearly equal to $S_l(q;P,T)$ and $S_g(q;P,T)$, correspondingly. Hence the order parameter vs temperature or pressure can be found using Eq. (90),

$$c_l(P,T) \approx \frac{S(q;P,T) - S_g(q;P,T)}{S_l(q;P,T) - S_g(q;P,T)} = \frac{S(q_m;P,T) - S_g(q_m;P,T)}{S_l(q_m;P,T) - S_g(q_m;P,T)}, \tag{91}$$

where $q_m$ is the position of the first maximum of $S_l(q;P,T)$. It is chosen as a convenient indicative value of $q$ taking into account that $c_l(P,T)$ does not depend on $q$. The accuracy of the $c_l(P,T)$ determination can be improved if one substitutes the integrated quantities $\Sigma(P,T)$, $\Sigma_l(P,T)$ and $\Sigma_g(P,T)$,

$$\Sigma(P,T) = \int_{q_m-\delta_1}^{q_m+\delta_1} S(q;P,T)dq \text{ , etc ,} \tag{92}$$

instead of $S(q;P,T)$, $S_l(q;P,T)$ and $S_g(q;P,T)$ in (91). The half-width of the first peak of the $S_l(q;P,T)$ is a proper value of the parameter $\delta_1$ determining the integration range in (92).

As it follows from Eq. (25), on the continued phase coexistence curve $c_l(P_e,T_e) = 0.5 + \varsigma$. Therefore at $P = P_e$ and $T = T_e$

$$S(q) = 0.5\left[S_l(q) + S_g(q)\right] + \varsigma\left[S_l(q) - S_g(q)\right] \tag{93}$$

Substitution of the extrapolated values of the $S_l(q;P_e,T_e)$ and $S_g(q;P_e,T_e)$ into Eq. (90) gives the extrapolated value $S_{ext}(q;P_e,T_e)$. Comparing it with the experimentally measured value, $S_{exp}(q;P_e,T_e)$, one can estimate the mesoscopic structural proximity effect.

To find the $c_l(P,T)$ of the supercritical water Novikov applied the formula like (90) to the measured double differential cross-section of the slow neutrons scattering.[37] Further investigations of the short-range structure factors and dynamics of the supercritical fluids are desirable.

## IX. CONCLUSIONS

The mesoscopic HPFM is formulated to describe the critical gas-liquid heterophase fluctuations. The mesoscopic description of the fluid is based on the "quantization" of the heterophase fluctuations. It allows considering the liquid-like and gas-like fractions as those consisting of the mesoscopic entities, *l*- and *g*-fluctuons.

The two-component order parameter of the heterophase fluid, which describes the amounts of the fluid-like and gas-like fractions, is nonlinearly connected with the density. Due to this nonlinearity, the effective Hamiltonian of the system loses the symmetry. The introduced asymmetry parameter $\varsigma$ is proportional to the difference between the mean numbers of molecules belonging to *l*- and *g*-fluctuons. It interplays with another gas-liquid asymmetry parameter $\varsigma_v$ (it is proportional to the difference between the mean densities of the liquid-like and gas-like fractions) which does not violate the effective Hamiltonian symmetry by itself but essentially impacts the asymmetry effects.

The estimation of the temperature range width of the fluctuation region, based on the OZ equation, shows that it is ~ $10^{-2}$ $T_c$ and it does not depend on the value of the nonuniversal mesoscopic correlation length $R_f$. This estimation augments the Levanyuk-Ginzburg crossover criterion.

The mesoscopic part of the structure factor of a fluid is the weighted sum of the short-range structure factors of the liquid-like and gas-like fractions. It is expressed in terms of the direct correlation functions of the *l*- and *g*-fluctuons. Thus the universal long-range density-density correlation function is augmented by the nonuniversal short-range constituent.

The asymmetry in the fluid criticality is a direct consequence of the effective Hamiltonian asymmetry. Since the coefficients $D_{2\beta}$, $D_{1-\alpha}$, and the Yang-Yang ratio $Y_R$ are scaled by the mesoscopic asymmetry parameters $\varsigma$ and $\varsigma_v$, one can conclude that the mesoscopic heterogeneities are responsible for the asymmetry in the fluid criticality.

On the phase diagrams the percolation lines of bubbles within a liquid phase and droplets within a gas phase bound the domains of the heterophase states of different topologies. They also determine the boundaries of the domains of applicability of the droplet model.

**AKNOWLEGMENT**

The work was partially supported by the Science and Technology Center in Ukraine, project #P45.

**APPENDIX A: SOME SOLUTIONS OF THE EQUATION OF STATE**

Using the standard normalized temperature and pressure, $\tau$ and $\pi^{\text{e}}$, the chemical potential (23) can be presented as an expansion in series of $\alpha$ degrees,

$$\frac{1}{T_c}\mu(P,T,\alpha) = \frac{1}{T_c}\mu(P,T,0) - \frac{1}{T_c}\left\{\frac{4-\varsigma^2}{4}\Delta\mu(\pi,\tau)\right\}\alpha(1-\varsigma\alpha) + \frac{1}{k}\left[\varsigma(\ln 2 - 1)\alpha - 2\tau\alpha^2 + 2\varsigma\tau\alpha^3 + \frac{4}{3}\alpha^4 - \frac{4}{3}\varsigma\alpha^5 + O(\tau\alpha^4) + ...\right]$$
(A.1)

Here

---

[e] The use of the conventional sign $\pi$ has not to lead to misunderstanding.

$$\Delta\mu(\pi,\tau) = \mu_g(\pi,\tau) - \mu_l(\pi,\tau); \quad T_c = g_2/2; \quad \pi = \frac{P-P_c}{P_c}; \quad \tau = \frac{T-T_c}{T_c}, \tag{A.2}$$

where $\mu_l(\pi,\tau)$ and $\mu_g(\pi,\tau)$ are the chemical potentials of the "pure" liquid and gaseous phases without heterophase fluctuations. The term $\sim \Delta\mu(\pi,\tau)\varsigma\alpha^2$, which is conserved in (A.1), violates the Hamiltonian symmetry.

The equation of state resulting from (A.1) and (26) is as follows,

$$-4\tau\alpha + 6\varsigma\tau\alpha^2 + \frac{16}{3}\alpha^3 = H(\pi,\tau)(1-2\varsigma\alpha) + O(\alpha^3). \tag{A.3}$$

Here

$$h(\pi,\tau) = -\frac{1}{T_c}\frac{\partial\mu(\pi,\tau)}{\partial\alpha}\bigg|_{\alpha=0} = -\frac{4-\varsigma^2}{4T_c}\Delta\mu(\pi,\tau) + \frac{\varsigma\ln 2}{2\bar{k}} \tag{A.4}$$

$$H(\pi,\tau) = \frac{1}{T_c}\bar{k}h(\pi,\tau); \tag{A.5}$$

is the "external field".

The last term in the right-hand side comes from the fluctuon mixing entropy. It leads to a shift of the phase coexistence temperature. Since $\varsigma/2\bar{k} \ll 1$, this term is insignificant and can be ignored. As Eq. (A.4) shows, the external field strength decreases with the asymmetry parameter $\varsigma$ increase.

The phase coexistence curve on the ($\pi,\tau$)-plane, $\pi = \pi_e(\tau)$, is determined by the equation

$$h(\pi,\tau) = 0, \tag{A.6}$$

which is analogous to Eq.(7). In linear approximation it is as follows,

$$\pi = \pi'_e \tau. \tag{A.7}$$

Here

$$\pi'_e = \frac{\partial h(\pi,\tau)}{\partial \tau} \bigg/ \frac{\partial h(\pi,\tau)}{\partial \pi}\bigg|_{\pi=\tau=0} \tag{A.8}$$

At constant temperature in the approximation linear in $\pi$ and $\tau$

$$h(\pi,\tau) = \frac{4-\varsigma^2}{4}\left[\Delta v_{gl}(\pi - \pi'_e \tau)\right]; \quad \Delta v_{gl} = v_g - v_l \tag{A.9}$$

At constant pressure

$$h(\pi,\tau) = \frac{4-\varsigma^2}{4}T_c\left[\Delta s_{gl}(\tau - \tau'_e \pi)\right]; \tag{A.10}$$

$$\tau'_e = (\pi'_e)^{-1}; \quad \Delta s_{gl} = s_g - s_l \tag{A.10a}$$

It follows from Eqs. (A.5), (A.9), (A.10) that at constant temperature

$$H(\pi,\tau) = \mathrm{M}_\pi \left[\pi - \pi_e(\tau)\right] \tag{A.11}$$

$$\mathrm{M}_\pi = \bar{k}\frac{P_c v_c}{T_c}\frac{4-\varsigma^2}{4}\frac{\Delta v_{gl}}{v_c} = \bar{k}\frac{P_c v_c}{T_c}\frac{\left(4-\varsigma^2\right)\varsigma_v}{4}\frac{\bar{v}}{v_c} = \bar{k}D_1\frac{P_c v_c}{T_c} \tag{A.11a}$$

At constant pressure

$$H(\pi,\tau) = \frac{1}{T_c}\bar{k}h(\pi,\tau) = \mathrm{M}_\tau(\tau - \tau'_e, \pi) \quad . \tag{A.12}$$

$$\mathrm{M}_\tau = \frac{4-\varsigma^2}{4}\bar{k}\Delta s_{gl} \tag{A.12a}$$

The solutions of the equation of state of type (A.3) are well-known. Let us represent some of them which are used in this paper. At $\tau < 0$ and $H(\pi,\tau) = 0$ there are three real solutions,

$$\alpha_{1,2}(\tau) = -\frac{9\varsigma}{16}\tau \pm \sqrt{-\frac{3}{4}\tau} + O\left(\tau^{3/2}\right); \quad \alpha_3(\tau) = 0 \tag{A.13}$$

The solutions $\alpha_{1,2}(\tau)$ are stable and the solution $\alpha_3(\tau)$ is unstable.

At $\tau > 0$ and $H(\pi,\tau) \neq 0$ there is one real and stable solution. At constant temperature,

$$\alpha(\pi,\tau) = \frac{H(\pi,\tau)}{4\tau}\left[1 - \frac{4}{3\tau}\left(\frac{H(\pi,\tau)}{4\tau}\right)^2\right] + ....$$

$$= \frac{1}{2}\tanh\left(\frac{H(\pi,\tau)}{2\tau}\right) + \frac{1-\tau}{6\tau}\left(\frac{H(\pi,\tau)}{2\tau}\right)^3 + ... \tag{A.14}$$

It follows from (A.3) that at $\tau > 0$ the $\alpha(T)$ is a steadily increasing function of $H$. As a result, the pure liquid phase ($\alpha = 0.5$) is formed at $H \to \infty$ while the gaseous phase ($\alpha = -0.5$) appears at $H \to -\infty$. Let us notice that the first term in (A.13) describes a continuous sigmoidal evolution of the order parameter within the range [-0.5 < α < 0.5] at $-\infty < h < \infty$. It means that the sum of the rest terms in (A.13) vanishes asymptotically, at large $|H|$. On the other hand, it is negligibly small compared with the first term at $|H| \to 0$. For this reason the first term in (A.14) can be taken as a reasonable approximation of $\alpha(\pi,\tau)$ if its behavior at $|H| < 1$ is not a subject of a particular interest.

At $|\tau| \ll h^{2/3}$ the cubic term is much larger than other terms in the left-hand side of (A.3). In this case

$$\alpha(\pi,\tau) = \left(\frac{3H(\pi,\tau)}{16}\right)^{1/3} \tag{A.15}$$

## APPENDIX B: SPECIFIC VOLUME AND DENSITY OF THE HETEROPHASE STATES

The specific volume of a heterophase state, as it follows from (24), is

$$v = \frac{\partial \mu(P,T)}{\partial P} = c_l v_l + c_g v_g = \bar{v}\left[1 - \frac{\varsigma_v(\varsigma + 4\alpha)}{4(1+\varsigma\alpha)}\right] \quad \text{(B.1)}$$

$$\bar{v} = \frac{v_l + v_g}{2}; \quad \varsigma_v = \frac{v_g - v_l}{\bar{v}}; \quad 0 < \varsigma_v < 2;$$

Let us remind that in the critical and supercritical regions the values of specific volumes $v_l = v_l(P,T)$ and $v_g = v_g(P,T)$ are the extrapolations of these quantities from the subcritical region. The term $O(\alpha^2)$ is conserved in (B.2) because it is $\sim \tau$ and its input in the ρ-diameter equation (see below (B.8)) is comparable with that of the linear term.

As it follows from (B.1), at the critical point

$$v(P_c,T_c) \equiv v_c = \frac{4 - \varsigma_v \varsigma}{4}\bar{v}(P_c,T_c) \quad \text{(B.2)}$$

Deducing the equation for the density, $\rho = v^{-1}$, it has to be taken into account that

$$\frac{\rho_l - \rho_g}{\bar{\rho}} = \varsigma_v; \quad \bar{\rho} = \frac{\rho_l + \rho_g}{2} \quad \text{(B.3)}$$

Hence, as it follows from (B.1),

$$v = \frac{4 - \varsigma\varsigma_v}{4}\left(1 - D_1\alpha + \varsigma D_1\alpha^2\right)\bar{v} + O(\alpha^3); \quad D_1 = \frac{\varsigma_v(4 - \varsigma^2)}{4 - \varsigma\varsigma_v} \quad \text{(B.4)}$$

$$\rho = \frac{4 - \varsigma_v^2}{4 - \varsigma\varsigma_v}\left[1 + D_1\alpha + D_2\alpha^2\right]\bar{\rho} + O(\alpha^3); \quad D_2 = D_1^2 - \varsigma D_1. \quad \text{(B.5)}$$

At the critical point the density is equal to

$$\rho_c = \frac{4 - \varsigma_v^2}{4 - \varsigma\varsigma_v}\bar{\rho}(T_c,P_c) \quad \text{(B.6)}$$

The liquid-gas coexistence curves in $(\rho,T)$-variables, as it follows from Eqs. (B.4), (B.5), and (28), are the following

$$\rho_{1,2} = \frac{4 - \varsigma_v^2}{4 - \varsigma\varsigma_v}\left[1 + D_1\alpha_{1,2} + \left(D_1^2 - \varsigma D_1\right)\alpha_{1,2}^2\right]\bar{\rho} \quad \text{(B.7)}$$

Therefore, the equation for the classic ρ-diameter is as follows,

$$\rho_d(\tau) = \frac{1}{2}[\rho_1(\tau) + \rho_2(\tau)] = \frac{4-\varsigma_v^2}{4-\varsigma_v\varsigma}\left[\frac{2+D_1(\alpha_1+\alpha_2)+(D_1^2-\varsigma D_1)(\alpha_1^2+\alpha_2^2)}{2}\right]\bar{\rho}(\tau)$$

$$\approx \left[1+\left(\frac{\bar{\rho}'}{\bar{\rho}(0)}+\frac{3}{16}D_1\varsigma-\frac{3}{4}D_2\right)\tau\right]\rho_c \quad (B.8)$$

Here

$$\rho_c = \frac{4-\varsigma_v^2}{4-\varsigma\varsigma_v}\bar{\rho}(T_c, P_c); \quad \bar{\rho}' = \frac{\partial \bar{\rho}(\tau)}{\partial \tau} \quad (B.9)$$

In the fluctuational region the coefficients $D_{2\beta}$ and $D_{1-\alpha}$ of the polynomial representation of the ρ-diameter (56),

$$\frac{\rho_d(\tilde{\tau})-\rho_c}{\rho_c} = \left(\frac{\bar{\rho}'}{\bar{\rho}(0)}+\frac{3}{16}\varsigma D_1\right)\tilde{\tau}+D_{2\beta}\tilde{\tau}^{2\beta}+D_{1-\alpha}\tilde{\tau}^{1-\alpha}+\ldots, \quad (B.10)$$

can be restored using the scaling analysis.[8,15-18] To use it, the Hamiltonian in the form of the Landau expansion has to be presented in terms of the order parameter

$$\eta = \frac{\rho-\rho_c}{\rho_c} \quad (B.11)$$

$D_{2\beta}$ and $D_{1-\alpha}$ are related to the coefficients in the terms ~ $\tau\eta^2, \tau\eta^3, \eta^4, \eta^5$ and $h\eta^2$. Keeping only these summands in the Landau expansion one has

$$\frac{\bar{k}}{T_c}\mu(P,T,\eta) = \frac{1}{2!}c_{21}\tau\eta^2+\frac{1}{3!}c_{31}\tau\eta^3+\frac{1}{4!}c_{40}\eta^4+\frac{1}{5!}c_{50}\eta^5+c_h h\eta^2+\ldots \quad (B.12)$$

The above-mentioned relations of the coefficients obtained from the scaling analyses are the following,

$$D_{2\beta} = Y_R \frac{6c_{21}}{c_{40}}, \quad (B.13)$$

$$D_{1-\alpha} = -\frac{3}{c_{40}}\left[\frac{c_{31}}{c_{21}}-\frac{c_{50}}{5c_{40}}\right]+\frac{2c_h(2-\alpha)}{c_{21}}, \quad (B.14)$$

$$Y_R = \frac{2c_{31}}{3c_{21}}-\frac{c_{50}}{5c_{40}}. \quad (B.15)$$

where $Y_R$ is the Yang-Yang ratio.[31]

As can be seen, the deviation of the ρ-diameter from linearity appears when the coefficients at the terms violating the Hamiltonian symmetry, $c_{31}$, $c_{50}$ and $c_h$, are non-zero.

Deriving the series expansion of $\eta(\alpha)$ from (B.5) and (B.11),

$$\eta(\alpha) = \frac{\rho(\alpha)-\rho_c}{\rho_c} = (D_1\alpha+D_2\alpha^2)\frac{\bar{\rho}}{\rho_c}+O(\alpha^3), \quad (B.16)$$

one can transform it into the expansion in series of $\alpha(\eta)$,

$$\alpha = \frac{1}{D_1}\eta - \frac{D_2}{D_1^2}\eta^2 + ... \tag{B.17}$$

The substitution of (B.17) in (A.1) gives the following values for the Hamiltonian coefficients (B.11)

$$c_{21} = \frac{4}{D_1^2}; \quad c_{31} = -\frac{12}{D_1^3}\frac{2D_2 + D_1\varsigma}{D_1}; \quad c_{40} = \frac{32}{D_1^4},$$

$$c_{50} = -\frac{120}{D_1^5}\frac{4D_2 + D_1\varsigma}{D_1}; \quad c_h = -\frac{\varsigma}{D_1^2} \tag{B.18}$$

Combining (B.13)-(B.15) and (B.18) one has

$$D_{2\beta} = -\frac{3}{4}D_1\varsigma, \tag{B.19}$$

$$D_{1-\alpha} = \frac{3}{4}(D_2 + D_1\varsigma) + 0.95\varsigma, \tag{B.20}$$

$$Y_R = -\frac{\varsigma}{D_1}. \tag{B.21}$$

**APPENDIX C: THE HPFM PARAMETERS**

The HPFM contains the set of the phenomenological parameters $g_2$, $\bar{k}$, $\varsigma$, $\varsigma_v$. These parameters determine the values of the coefficients $D_1$, $D_2$, $Y_R$, the strength of the external field $H(P,T)$ and other physical quantities. The estimation of the ranges of plausible values for the parameters on the basis of experimental data allows to estimate the values of the coefficients determining the measurable physical quantities.

The thermodynamic quantities, $\mu_l(P,T)$, $\mu_g(P,T)$ and their derivatives are regarded as those insignificantly differing from the corresponding homophase quantities extrapolated from the subcritical into the critical and supercritical region. The validity of the extrapolated quantities can be verified by comparing the theoretical predictions (e.g. relations (63-68)) and the experimental data.

Since the equation of state in the HPFM contains more than two free parameters, it can not be reduced to the equation of corresponding states with numerical coefficients, like the van der Waals equation. At the same time, due to the success of the van der Waals model, one can suspect that the parameters of the HPFM model vary within comparatively narrow ranges. We have found that $g_2 = 2T_c$, $\bar{k} >> 1$, $|\varsigma| < 2$ and $0 < \varsigma_v < 2$. To estimate the ranges of plausible values of the parameters $\varsigma$ and $\varsigma_v$, let us compare the formula (55) for the ρ-diameter with the experimental data. It was found[43] that for eight atomic and molecular fluids with good accuracy

$$\rho_d(\tau) = \rho_c(1 - 0.75\tau) \qquad (C.1)$$

Equating the right sides of Eqs. (55) and (C.1) we obtain

$$\tilde{\rho}' = \frac{\overline{\rho}'}{\overline{\rho}(0)} = D_1\left(\frac{3}{4}D_1 - \frac{19}{16}\varsigma\right) - 0.75 \, ; \; D_1 = \frac{\varsigma_v(4-\varsigma^2)}{4-\varsigma\varsigma_v}. \qquad (C.2)$$

Since the thermal expansion coefficient of a fluid is positive, we have to demand the fulfillment of the condition

$$\tilde{\rho}' < 0 \qquad (C.3)$$

The graphic representation of $D_1$ and $\tilde{\rho}'$ vs $\varsigma$ at fixed values of $\varsigma_v$ are shown in Fig. 10. As it is seen, the condition (C.3) can be fulfilled if $\varsigma_v < 0.8$.

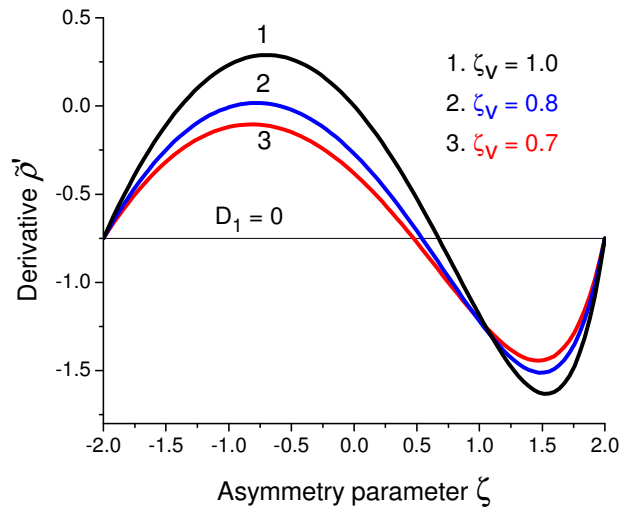

Fig. 10. The derivative $\tilde{\rho}'$ vs the parameter $\varsigma$ as it follows from Eq. (C.2).

To estimate the bounds of the range of parameter $\varsigma$ values, let us take into account that the experimental and simulation data show that the modulus of Yang-Yang ratio, $|Y_R|$, as a rule does not exceed 0.5.[15,18] Taking $|Y_R| < |Y_R|_{max} \leq 0.5$, from Eq. (60) we have

$$|\varsigma| \leq \varsigma_v |Y_R|_{max} \leq 0.4 \qquad (C.4)$$

Since $\varsigma_v < 0.8$ and $D_1 \leq \varsigma_v$ one can conclude that $|\varsigma| < 0.8$. Apparently $|\varsigma| < 0.4$.

**APPENDIX D: LOCI OF THE DPL AND BPL ON THE PHASE DIAGRAMS**

**1. The (ρ,T)-diagram**

The DPL and BPL on the phase diagrams are the loci of the percolation thresholds points. Let us introduce the definition of the DPL and BPL on the $(\rho,\tau)$-plane. Using Eq. (78) let us determine the density values at the percolation thresholds. The normalized density linear in $\tau$, $\rho_n(\tau)$, follows from Eq.(55),

$$\rho_n(\tau) = \frac{\rho(\tau) - \rho_c}{\rho_c} = D_1\alpha + D_2\alpha^2 + \left[1 + D_1\alpha + D_2\alpha^2\right]\tilde{\rho}'\tau \tag{D.1}$$

By substituting the values $\alpha_d$ and $\alpha_b$ (which are determined by Eq. (78)) in (D.1) one obtains the relations

$$\rho_{n,d} = D_1\alpha_d + D_2\alpha_d^2 + \left[1 + D_1\alpha_d + D_2\alpha_d^2\right]\tilde{\rho}'\tau \tag{D.2}$$

and

$$\rho_{n,b} = D_1\alpha_b + D_2\alpha_b^2 + \left[1 + D_1\alpha_b + D_2\alpha_b^2\right]\tilde{\rho}'\tau \tag{D.3}$$

Eqs. (D.1) and (D.2) determine the desirable loci of the DPL and BPL on the $(\rho,\tau)$-plane. It is seen that in this approximation they are strait lines with the slopes

$$\frac{\partial \rho_{n,d}}{\partial \tau} = \left[1 + D_1\alpha_d + D_2\alpha_d^2\right]\tilde{\rho}' \tag{D.4}$$

$$\frac{\partial \rho_{n,b}}{\partial \tau} = \left[1 + D_1\alpha_b + D_2\alpha_b^2\right]\tilde{\rho}' \tag{D.5}$$

correspondingly.

The cross points of the DPL and BPL with the projection of the phase coexistence curve on the $(\rho,\tau)$-plane are the end points bounding these lines. To find their coordinates let us use the gas-liquid coexistence curves experimentally determined by Guggenheim[43]. They can be properly fitted by a cubic parabola,

$$\rho_{n,\text{liquid}} = 1.8(-\tau)^{1/3} \tag{D.6}$$

$$\rho_{n,\text{gas}} = -1.8(-\tau)^{1/3} \tag{D.7}$$

The desirable coordinates of the cross points are correspondingly equal to

$$\rho_{n,d} = D_1\alpha_d + D_2\alpha_d^2; \quad \tau_d = \left(\frac{D_1\alpha_d + D_2\alpha_d^2}{1.8}\right)^3 \tag{D.8}$$

and

$$\rho_{n,b} = D_1\alpha_b + D_2\alpha_b^2 \ ; \quad \tau_b = -\left(\frac{D_1\alpha_b + D_2\alpha_b^2}{1.8}\right)^3 \tag{D.9}$$

Since $|\alpha_d|, \alpha_b < 0.5$ and $D_1 < 1$, the values $|\tau_d|$ and $|\tau_b|$ as a rule do not exceed $10^{-2}$ at plausible values of the parameters $\varsigma$ and $\varsigma_v$, i.e. at $\varsigma < 0.4$ and $\varsigma_v < 0.8$ (see Appendix C). For example, taking $\varsigma = 0.2$ and $\varsigma_v = 0.8$, we have $\tau_d \approx 0.9*10^{-3}$ and $\tau_b \approx -5*10^{-3}$.

The graphic representation of the DPL and BPL loci on the $(\rho,\tau)$-phase diagram is shown in Fig. 7.

## 2. The (P,T)-diagram

The DPL on the $(\pi,\tau)$-phase diagram is the locus of points $\alpha_d(\pi,\tau)$. The substitution of $\alpha_d$ into the equation of state (30) leads to the desired solution in a classical approximation,

$$4\alpha_b\tau - 6\varsigma\tau\alpha_b^2 + \frac{16}{3}\alpha_b^3 = H(\pi,\tau) \tag{D.10}$$

At fixed $\tau$ the approximation the $H(\pi,\tau)$ linear in $\pi$ is determined by Eqs. (A.11), (B.2)

$$H(\pi,\tau) = M_\pi [\pi_b - \pi'_e \tau]; \tag{D.11}$$

$$M_\pi = \frac{(4-\varsigma^2)\varsigma_v}{4-\varsigma\varsigma_v}\frac{P_c v_c}{T_c} = \bar{k} D_1 \frac{P_c v_c}{T_c} \ . \tag{D.12}$$

As it follows from (D.1)-(D.3), the PBL locus is

$$\pi_b(\tau) = \frac{1}{M_\pi}\left[\frac{16}{3}\alpha_b^3 + \left(M_\pi\pi'_e + 4\alpha_b - 6\varsigma\alpha_b^2\right)\tau\right] \qquad \tau \geq 0.05 \tag{D.13}$$

The estimations of the fluctuation region width obtained in Sect.4 show that the classic model and, consequently, the solution (D.13) is valid at $|\tau| \geq 0.05$.

Replacing $\alpha_b$ by $\alpha_d$ in (D.13) we get the analogous equation for the DPL,

$$\pi_d(\tau) = \frac{1}{M_\pi}\left[\frac{16}{3}\alpha_d^3 + \left(M_\pi\pi'_e + 4\alpha_d - 6\varsigma\alpha_d^2\right)\tau\right]; \quad \tau \geq 0.05 \tag{D.14}$$

The behavior of $\pi_b(\tau)$ at smaller values of $\tau < 0.05$ is unknown, but as it is found while analyzing the percolation lines loci on the $(\rho,\tau)$-phase diagram, the $\tau$-coordinates of the cross points of DPL and PBL with the phase coexistence curve are determined by Eqs. (D.8), (D.9).

Let us estimate the parameter $M_\pi$ determined by Eq.(A.11). The measurements show that the values of the dimensionless parameter $P_c v_c / T_c$ of many fluids are in the range [0.23–0.3] (e.g.

[11]). In the van der Waals model it is equal to 0.375. Taking into account that $D_1 < 0.8$, one can conclude that $\mathrm{M}_\pi$ is the quantity ~ 1. For example, with $D_1 = 0.7$ and $\bar{k} = 12$ $\mathrm{M}_\pi \in [2 - 3.2]$.

The parameter $\pi'_e$ usually varies within the range [3-5]. For example, for argon $\pi'_e \approx 3.5$, for water $\pi'_e \approx 4$.

The graphic representation of the DPL and BPL loci on the $(\pi, \tau)$-phase diagram is shown in Fig. 8.